\documentclass[fleqn,usenatbib]{mnras}

\usepackage{newtxtext,newtxmath}

\usepackage[T1]{fontenc}
\usepackage{float}
\usepackage{ae,aecompl}
\usepackage{lscape}
\usepackage{multirow}
\usepackage{threeparttable}

\usepackage{graphicx}	
\usepackage{amsmath}	
\usepackage{longtable}
\newcommand{\logg}{log $\it{g}$}
\newcommand{\cemps}{CEMP-$\it{s}$}
\newcommand{\cemprs}{CEMP-$\it{r/s}$}
\newcommand{\cempr}{CEMP-$\it{r}$}
\newcommand{\cempno}{CEMP-$\it{no}$}
\newcommand{\iprocess}{$\it{i}$-process}
\newcommand{\sprocess}{$\it{s}$-process}
\newcommand{\rprocess}{$\it{r}$-process}

\title[Abundances in CEMP stars]{Optical and NIR spectroscopy of cool CEMP stars to probe the nucleosynthesis in low mass AGB binary system}

\author[A. Susmitha et al.]{
A. Susmitha$^{1}$\thanks{E-mail: susmitha.antony@tifr.res.in (AS)},
D.K. Ojha$^{1}$,
T. Sivarani$^{2}$,
J.P. Ninan$^{3}$,
A. Bandyopadhyay$^{4}$, \\
\newauthor {Arun Surya $^{5}$, Athira Unni $^{2}$}
\\
$^{1}$Tata Institute of Fundamental Research, Colaba, Mumbai- 400005, India\\
$^{2}$Indian Institute of Astrophysics, Koramangala II block-560034, Bengaluru, India\\
$^{3}$ Dept. Of Astronomy and Astrophysics, 525 Davey Lab, Pennsylvania State University, University Park, 16802,
The United States.\\
$^{4}$ Aryabhatta Research Institute of Observational Sciences, Manora Peak, Nainital-263001,  India \\
$^{5}$ Center for Astrophysics And Space Sciences, University of California, San Diego, 92093, The United States \\
}

\date{Accepted XXX. Received YYY; in original form ZZZ}

\pubyear{2015}

\begin{document}
\label{firstpage}
\pagerange{\pageref{firstpage}--\pageref{lastpage}}
\maketitle

\begin{abstract}
We present the abundance analyses of 7 Carbon enhanced metal-poor (CEMP) stars to understand the origin of carbon in them. We used high-resolution optical spectra to derive abundances of various elements. We also used low-resolution Near-Infrared (NIR) spectra to derive the abundance of O and $^{12}$C/$^{13}$C from the CO molecular band and compared their values with those derived  from high-resolution optical spectra. We  identified a good agreement between the values. Thus, in cool CEMP stars, the NIR observations complement the high-resolution optical observations to derive the oxygen abundance and the $^{12}$C/$^{13}$C ratio. This enables us to probe fainter cool CEMP stars using NIR spectroscopy. C, N, O abundances of all the program stars in this study show abundances that are consistent with binary mass transfer from a  low-mass low-metallicity Asymptotic Giant Branch (AGB) companion which is further supported by the presence of enhancement in neutron-capture elements and detection of radial velocity variation.  One of the stars show abundance patterns similar to a \cemps\ star whereas the abundance pattern of the rest of the stars satisfy the criteria required to classify them as \cemprs\ stars. The sub-classification of some of the stars studied here is revisited. The abundance of neutron-capture elements in these \cemprs\ stars resembles to that of \iprocess\ models where proton ingestion episodes in the companion low-mass low-metallicity AGB stars produce the necessary neutron density required for the onset of \iprocess.
 
\end{abstract}




\section{Introduction}
Spectroscopic studies of metal-poor stars identified from early large surveys like Hamburg/ESO survey and HK survey brought out the fact that a significant fraction of
Galactic halo stars  exhibit strong enhancement in  carbon  with [C/Fe] $>$+0.7 \footnote{The abundance of one element with respect to another is expressed  by the comparison of their ratios with respect to the sun and is indicated as, [A/B] = log$_{10}(N_{A}/N_{B})_{*} - log_{10}(N_{A}/N_{B})_{\odot} $, where N$_{A}$ and N$_{B}$ are the number densities of elements A and B and * and $\odot$ denote the stellar and solar values.}\citep{beers1992AJ, christlieb2001aa,aoki2007apj, yanny2009aj}.
The fraction of so called carbon enhanced metal-poor [CEMP] stars increases with
decreasing metallicity \citep{lucatello2005ApJ,lucatello2006apj,aoki2007apj,aoki2013AJ,Yong2013a,placco2014apj}
 and it is of great interest due to their high carbon abundance and the sizable fraction in the low-metallicity regime.
Various follow-up studies have revealed that these CEMP stars also show  peculiar  elemental abundance pattern along with the strong over abundance in C. Based on these peculiar abundance pattern, CEMP stars are broadly classified
into two categories \citep{beers-christlieb2005ARA&A,jonsell2006aa, aoki2007apj, masseron2010aa}; one shows over abundance of neutron capture elements including the elements from  slow neutron-capture process ($\it {s}$-process)  or/and rapid neutron-capture ($\it {r}$-process) process (\cemps,\cempr, \cemprs) and other does not
show any enhancement in neutron capture elements (\cempno). \\
Among the CEMP population, \cemps\ sub-class constitutes  around 80\% of the population in the metal-rich end \citep[based on their sample of $\sim$40 stars altogether]{aoki2007apj, hansentt2016aa2} and multi-epoch
observations of these stars reported that
majority of stars in the \cemps\ class are members
of binary systems \citep{lucatello2005ApJ, Starkenburg2014}. This favors the idea of mass transfer from a companion AGB star which is more massive compared to the primary and therefore would  have evolved to a cool white dwarf \citep[and references therein]{bisterzo2012MNRAS, abate2015}. 
 For the case of  \cempr\ stars, only  a few of them are known till date \citep{Sneden2003, hansentt2015apj} and the possible origin of \rprocess\ elements in this subclass include supernovae (SNe) nucleosynthesis \citep{wheeler1998apj, qian2007, sneden2008ARAA, winteler2012ApJ}  and/or neutron star merger events \citep{lattimer1974Apj, thieleman2017, watson2019Nature}. There are also \cemprs\ stars, that constitute relatively a larger fraction compared to \cempr\ stars, exhibiting enhancement in $\it {r}$- and $\it {s}$-process elements. They also show radial velocity variations, hence  two different sources for such  enrichment of $\it {s}$- and \rprocess\  elements are considered. Some hypothesize that
carbon and \sprocess\ abundance would have similar
origin as that of the \cemps\ stars \citep{sivarani2004aa, masseron2010aa, bisterzo2011} whereas the origin of \rprocess\ elements is considered to be from SNe or neutron-star mergers which pre-enrich the birth cloud of the binary stars with \rprocess\ elements \citep{bisterzo2012MNRAS}. This two-source pollution could not satisfactorily explain the observed number of \cemprs\ stars as well as the correlation between the enrichment in the $\it {s}$-process and \rprocess\ elements \citep{abate2015}.  Recently, nuclear network calculations using neutron densities halfway between $\it{s}$-and {\it r}-processes  were able to produce the abundance pattern similar to \cemprs\ stars \citep{hampel2016apj, denissenkov2017apj} thus invoking the possibility of a  single stellar site for the production of $\it{s}$- and \rprocess\ in \cemprs\ stars instead of two-source pollution.  In the literature sometimes \cemprs\ stars are referred as CEMP-$\it{i}$ stars because 
the intermediate neutron-capture process seems capable of reproducing the mixed $\it {s}$- and ${r}$- process abundance pattern. \\
While majority of the CEMP population exhibit the signatures of a  binary companion, \cempno\ stars are often found not to be associated with binaries \citep{hansentt2016aa}  thus they reflect the ISM abundance from which the star was born. Also, \cempno\ stars are  found to be dominating the CEMP population at lower metallicities \citep[]{salvadori2015mnras}. Models
of faint SNe that experienced mixing and fallback and models of zero-metallicity-spin stars having
high rotational velocity are some of the proposed scenarios for explaining the peculiar
abundance patterns shown by CEMP-$\it {no}$ stars \citep{Umeda2003, Meynet2006, Yoon2016, hansentt2016aa}. These progenitors expel huge amount of C,N and O to the ISM. \\
Despite various astrophysical processes are identified as
a source for carbon in \cemprs\ stars, the
masses of progenitors and multiple nucleosynthetic
processes or their sites to explain the diverse abundance pattern of the light and heavy elements are still unclear. \\
To fully understand the sites and mechanisms that contribute the enrichment of light and heavy elements in \cemprs\, we require measurements of  C, N and O for a large sample of CEMP stars along with other elemental abundances. The abundance of C and N can be determined from low-resolution optical spectra of CEMP stars using the
CH, C${_2}$ and CN features. But O abundance measurements  require high-resolution spectroscopy. Typical high-resolution oxygen abundances are made from the forbidden [OI] lines at $\lambda$6300, $\lambda$6363 and O I triplet near $\lambda$7774. 
However,  the O I  triplet lines at $\lambda$7774  are  strongly affected by non-local thermodynamic equilibrium (NLTE)  and local thermodynamic equilibrium (LTE) approximation results in over estimation of the abundances \citep{asplund2005, amarsi2016MNRAS, amarsi2019aa}. 
But the forbidden [OI] lines at $\lambda$6300 and $\lambda$6363  do not show any significant departures from LTE even though they do suffer from 3D effects. These 3D effects cause an abundance variation maximum up to 0.2 dex \citep{caffau2015aa, amarsi2016MNRAS}. Hence, more  reliable abundance measurements thus come from the  forbidden [O I] lines at 6300\AA\ and 6363\AA. 
But these lines are very weak at lower metallicities and it requires hours of exposures to detect the feature \citep{schuler2006AJ}.
\begin{table*}
\centering
\caption{The details of the program stars.}
\label{tab:obsdetail}
\begin{tabular}{ccccccccccc}

\hline
Star & RA & DEC & Vmag & Kmag & MJD & RV$_{helio}$ & SNR at & SNR at &  Reference: \\ 
&(J2000) & (J2000) & & & & (kms$^{-1}$) & 2.1 $\mu$m & 4800 \AA &  \\
\hline
HD 5223 & 00:54:13.61 & +24:04:01.52 & 8.5 & 5.7& 57775.68434693 & -242.9&66.3& 58.2 &1 \\
HE 0314-0143 & 03:17:22.18 & -01:32:36.51 & 12.6 & 8.0& 57784.59031674 & -33.6 &45.3& 27.5 &1,2 \\
HE 1152-0355 & 11:55:06.06 &-04:12:24.71 & 11.4  &8.4& 57801.81004023 & 428.6 &36.2&  26.7&1,2\\
HE 0017+0055 & 00:20:21.60 & +01:12:06.83 & 11.7 & 8.5 & 58720.81863073 & -83.2 &33.8& 56.1 &1,2 \\
BD+42 2173 & 11:00:00.44 & +41:36:07.50  & 8.1 & 7.4 & 58575.57389648 & -59.9 &61.6& 50.2&1 \\
 HD 187216 & 19:24: 18.35 & +85:21:57.19 & 9.6 & 6.0 & 58575.85302417 & -122.4&44.7& 51.6 &1\\
 HE 1418+0150 & 14:21:01.15 & +01:37:17.76 & 12.2  & 9.1 & 57801.97132139 & -22.4 &44.6&  26.2&1,2\\
\hline
\end{tabular}
\begin{tablenotes}
\item References: 1: \citet{masseron2010aa}; 2: \citet{kennedy2011aj} and references therein. 
\end{tablenotes}
\end{table*} 
Many cool CEMP stars are unexplored due to the complexity of analysis and need for high resolution high S/N data. But the near-infrared (NIR) first-overtone ro-vibrational bands of CO at 2.3 $\mu$m gives a possible alternate to this problem
because of their detection  even at lower resolution \citep{beers-sivarani2007aj, kennedy2011aj,hansen2019aa}. In CEMP stars, the C/O ratio is $>$ 1.0 i.e; the carbon abundance exceeds the oxygen abundance.
So essentially all the oxygen would be  locked up in  CO molecules and these NIR lines will provide a sensitive
probe for the oxygen abundance in CEMP stars. \\
 The carbon isotopic ratio $^{12}$C/$^{13}$C  is an important parameter to understand the degree of mixing in the progenitors or the observed star itself \citep{spite2005, Spite2006A&A,sivarani2006aa}  and is usually derived from the  optical spectra using the $^{12}$C$_{2}$
band head at 4737 \AA\ and the $^{12}$C$^{13}$C band head at 4744\AA\ . But in cool stars, the $^{12}$C$_{2}$
band head at 4737 \AA\ may be saturated and leads to a large derived $^{12}$C/$^{13}$C ratio.
This problem can be resolved by using the CO features at 2.3 $\mu$m where the   $^{13}$CO lines are well separated  from the $^{12}$CO lines. So it  provides an accurate and better
estimates of the mixing diagnostic $^{12}$C/$^{13}$C ratio than deriving it  from the optical spectra \citep{beers-sivarani2007aj}, for cool CEMP stars.\\
In this paper, observations of 7 CEMP stars are reported to understand the origin of various elements in them and to validate
the oxygen abundances derived from NIR spectra with that of [OI]6300 line in optical.
The paper is organized as follows: In section 2,
sample selection and observation have been mentioned. Section 3 and section 4  describe the details of stellar parameters 
and abundance analysis. In section 5, the results are discussed with how the companion's properties can be constrained with the abundance pattern of  light and heavy elements  and in section 6, the paper is concluded.
\section{Observation and data reduction}
One of the motivations of this study is to measure the C, N and O abundances, and compare the oxygen 
abundance from   CO features in the NIR regime and oxygen from [OI] at 6300 \AA.
So the target selection considered were, metal poor stars that are enhanced in carbon abundance, low effective temperature (T$_{eff} <$ 5000K) and Vmag < 13, suitable for observations using a 2m class telescope. At such  cool temperatures and with C/O $>$ 1, all the oxygen will be locked up as CO molecule.
So the program stars were chosen from literature where the carbon has been previously measured which has low effective temperature.
The details of the targets are given in the table \ref{tab:obsdetail}. \\  
The targets were observed using the TIFR (Tata Institute of Fundamental Research) Near Infrared Spectrometer and Imager (TIRSPEC) \citep{Ninan2014jai} and Hanle Echelle Spectrograph  (HESP) \citep{sriram2018SPIE} mounted on 2m Himalayan {\it Chandra} Telescope (HCT), Hanle. TIRSPEC was used to capture the 
CO bands at 2.3 $\mu$m and HESP was used for the optical observations. 
The NIR spectra have been taken in the cross-dispersed mode using a slit S3 (width is 1.97"). This setup covers  both
the  H and K bands   simultaneously, covering a wavelength of 1.50 - 1.84 $\mu$m \& 1.95 - 2.45 $\mu$m at a resolution R $\sim$ 800.
Dithering along the slit was performed 
for accurate background subtraction, we also avoided regions of bad pixels  by suitably choosing the target location on the slit. 
A0 type stars at the same air mass were  also observed along with the targets  to correct for the 
telluric lines in the spectra. In order to avoid the counts reaching the non-linearity regime of 
the detector, the observations were performed as multiple exposures of 500sec. The final spectrum thus obtained had a signal to noise ratio (SNR)  $\sim$ 30.
The NIR data reduction has been performed using semi automated pipeline developed by \citet{Ninan2014jai}
in which the dark correction, flat fielding, background subtraction, spectrum extraction, wavelength 
calibration, telluric line correction and continuum fitting were performed using codes in python
which made use of standard modules like astropy, numpy, matplotlib and PyRAF\footnote{PyRAF is a 
product of the Space Telescope Science Institute, which is operated by AURA for NASA}. \\
The optical observations of the program stars were taken in high-resolution (R = 30000) using  HESP. They were observed in multiple frames of 45 minute
exposures depending on the brightness and obtained the spectra in the star-sky mode 
which enabled the sky to be subtracted from the object fibre.
ThAr spectrum was taken along with each program star spectrum for the 
wavelength calibration. 
The spectral reduction included, trimming, bias subtraction, aperture extraction and wavelength
calibration.  These steps were performed using  PyRAF. 
The final normalized and co-added spectrum covers a wavelength range from 4000-10500 \AA.
\section{stellar parameters and chemical abundances}
The spectra  of all the program stars are dominated by 
molecular features because of high carbon abundance and low effective temperature.
This hindered us from getting  clean lines of iron (Fe) for estimating 
the stellar parameters using equivalent width method. 
So, we performed a full spectrum synthesis to derive abundances from  the atomic lines. \\
We used the spectral synthesis code TURBOSPECTRUM developed by \citet{Plez2012ascl} to derive abundances and obtain the stellar parameters. We used the stellar atmospheric models by  \citet{meszaros2012aj} in which the ATLAS9 and
MARCS codes were modified with an updated H$_{2}$O linelist and
with a wide range of carbon- and $\alpha$-element enhancements. Whenever required, we interpolated the models from the grid of model photospheres provided by \citet{meszaros2012aj} to obtain the intermediate parameter values. LTE has been assumed for all
species. But we referred  \citet{mashonkina1999aa, mashonkina2000aa, andrievsky2009aa, Andrievsky2010AA, Lind2011, Mashonkina2014} and  \citet{hansencj2020aa} for NLTE corrections of various elements and whenever available we applied their  corrections to the derived abundances.
We adopted the Solar abundances from \citet{asplund2009araa}
and Solar isotopic ratios were used unless otherwise specified. The line lists for atomic lines
were assembled from the Vienna Atomic Line Database (VALD)  database \citep{kupka1999aas} and the details of the same are mentioned in the table \ref{tab:linelist}. Hyperfine structure (HFS) has
been accounted for Li, Sc, Ba, La and Eu.
For the case of molecular linelists, we used the CH line list compiled by T.
Masseron (priv. comm.)  and CN data from \citet{plez-cohen2005aa}. The linelists for C$_{2}$ and NIR CO molecular features were taken from the Kurucz database\footnote{http://kurucz.harvard.edu/linelists/linesmol/}.
\subsection{Stellar parameters}
The effective temperatures of all the program stars were calculated from J-H, J-K and V-K colors using various color transformation relations \citep{Alonso1996,Alonso1999, ramirezMelindez2005ApJ}. For this, whenever available, we used the metallicity quoted in the literature as the initial metallicity  of the respective star and the reddening values, E(B-V), from \citet{Schlegel1998ApJ}. The derived photometric temperature values are consistent within 150K. Among these,  the temperature derived from the V-K color index is considered superior due to its large span in the wavelength of the bands involved.  
Also, these bands are relatively free from contamination from molecular C.
So we adopted the temperature derived from the V-K color as the star's temperature which is also  consistent within 150K with literature values. Hence, we adopt this value as uncertainty in the temperature measurements, for deriving abundances. 
We have also cross-verified the temperature of the star using the available data from {\it Gaia} data release 2\footnote{https://gea.esac.esa.int/archive/}. The available {\it Gaia} temperatures are matching within the error ($\pm$ 150 K). 
\\
The surface gravity,  \logg, was derived by fitting the wings of Mg I triplet around 5172 \AA\ using a synthetic grid of model spectra where we adopted the temperature derived from the above method as the model temperature value and for various values
of \logg\ with a step size of 0.25 dex. The best fit value was 
found by the goodness of the fit and chosen as the final \logg. While fixing \logg, we also changed the carbon abundance from [C/Fe]=0.0 to [C/Fe]= +1.5 to simultaneously fit the C$_{2}$ band-head at 5165 \AA\ to converge on to the best fit parameters. We have checked the effect of carbon abundance on the wings of Mg I lines by fitting the Mg I lines with different values of [C/Fe]. The effect is found to be negligible (refer figure \ref{fig:c2mg5172}). 
We identified that the \logg\ of the program stars ranges from 0.25 to 1.9. 
In order to understand the evolutionary phase of the program stars, we obtained the luminosity of the stars from {\it Gaia} DR2 data and placed the stars on their respective isochrones from BASTI stellar evolution database \citep{pietrinferni2004apj} and identified that all the program stars fall in the upper part of  the red giant branch (RGB). 
 The metallicities of the stars were derived by fitting the iron features in the visual
spectra  where the molecular features have less influence on the continuum  (e.g., Fe lines in  the range 5190 \AA\ - 6450 \AA). The best fit value was chosen as the final value with an error of $\pm$0.2 dex. The uncertainty was chosen in such a way that the best fit does not change while changing the Fe abundance.  
We derived the  microturbulence velocity ($\xi$) by simultaneously matching the iron abundance of  weak and strong FeI lines in the spectra. We fitted the unblended iron features by varying the $\xi$ by a step-size of 0.2 kms$^{-1}$ and the best fit value was taken as the final microturbulent velocity. 
The final adopted stellar parameters  with the uncertainty in measurements are listed in the table \ref{tab:stparam}
\begin{table}
\centering
\caption{The atmospheric parameters of the program stars from this study. The uncertainties in the measurements of stellar parameters is also mentioned in the table.}
\label{tab:stparam}
\begin{tabular}{|c|c|c|c|c|c|}
\hline
star name& T$_{eff}$& \logg & [Fe/H] & $\xi$ & type\\
        & K & cms$^{-2}$&dex& kms$^{-1}$ &\\
        & $\pm$150 & $\pm$0.25 & $\pm$0.2 & $\pm$ 0.2 & \\ 
\hline
\hline
HD 5223 & 4335 & 1.20 & -2.1 & 2.0 & r/s\\
HE 0314-0143 & 3881 & 0.25 & -1.9 & 2.1 & r/s \\
HE 1152-0355 & 4200& 0.25 & -1.7 & 2.2 & s\\
HE 0017+0055 & 4240 & 1.00 & -2.6 & 1.9 & r/s \\
HE 1418+0150 &  4150 & 1.80 & -2.0 & 2.0 & r/s\\
BD+42 2173  & 4430  & 1.90 & -1.6 & 1.5 & r/s\\
HD 187216      & 3920 & 0.80 &  -2.5 & 2.0 & r/s\\
\hline
\end{tabular}
\end{table}
\section{Abundances}
The abundances of key elements such as Li, C, N, O, Mg, Na, Ca, Y, Ba, Eu and Sm  were derived using the spectral synthesis method where the input model parameters were chosen from the table \ref{tab:stparam} . Using these model parameters, the  synthetic spectral grid was generated by varying the input elemental abundances in steps of 0.2 dex. This synthetic spectra were used to fit the respective spectral line and the best fit value was chosen as the final abundance. The derived abundances are quoted in the table \ref{tab:abund-chem}. The details of the abundance measurements for each element  are briefly described below.
\subsection{Uncertainties in the abundance measurements} \label{error}
The stellar parameters and the elemental abundances are estimated using spectral synthesis method. So, the uncertainty in the abundance values was estimated through the goodness of the least squares fit and  are mentioned in the  table \ref{tab:syserror}. \\
The systematic errors in our abundances were derived using models where each stellar parameter (T$_{eff}$, \logg, $\xi$)  
was varied by a fixed amount ($\pm$150 K, $\pm$0.25 dex, $\pm$0.2 km s$^{-1}$) while keeping other parameters constant 
and from those, new abundances were computed. To perform the analysis, we choose HD 5223  as the representative star for the program stars in this study.  The resulting  change on abundance variations upon these
parameter variations as well as their combined effect,  calculated by adding them in  quadrature, are listed in the table \ref{tab:syserror}.
\begin{table*}
 \caption{The [X/Fe] of various elements are reported in the table. The corresponding 1$\sigma$ uncertainty in the abundance measurements, $\sigma_{log \epsilon(X)}$,  is  mentioned in the last column.}
    \centering
    \begin{tabular}{cccccccccc}
    \hline
    \hline
    Element& Solar & HD 5223 & HE 1152-0355 & HE 1418+0150 & BD+42 2173 &  HD 187216 & HE 0017+0055 & HE 0314-0143 & $\sigma_{log \epsilon(X)}$   \\
    \hline
    \hline
Li$^{\S}$  	& 1.05	&  $<$0.00	& $<$0.00	&  $<$-0.05	&  $<$0.00	&	  $<$0.00	&  $<$-0.10	&$<$-0.50 & 0.30  \\
C   	& 8.43	& 1.35	& 1.35	& 1.30	& 1.10	& 	 1.45	& 2.10	& 0.95 & 0.20 \\
N   	& 7.83	& 0.73	& 0.95	& 1.20	& 1.00	& 	 0.93	& 1.65	& 0.55 & 0.20  \\
O    	& 8.69	& 0.35	& 0.50	& 1.00	& 0.35	&	 0.05	& 0.90	& 1.02 & 0.30  \\
Na (LTE)  	& 6.24	& 0.20	& -0.10	&0.60	&  0.80	&	 0.50	& 0.70	& 0.30 & 0.20 \\
Na (NLTE) 	& 6.24	& 0.13$^{*}$	&---	& 0.45$^{*}$	& 0.60$^{*}$	&---	& 0.61$^{*}$	& 0.20$^{*}$ & 0.20  \\
Al  	& 6.45	&$<$ -0.00	&$<$-0.20	&$<$-0.40	& 0.65	& 	$<$-0.10	& $<$0.00	&$<$-0.00 & 0.30 	\\
Mg  	& 7.60	&-0.10	& 0.40	& 0.10	& 0.40	& 	 0.30	& 0.00	& 0.40	& 0.20 \\
K(LTE)   	& 5.03	&...	& 0.70	& 0.10	& 0.80	& 	 0.10	& 0.60	& 0.00 & 0.20	\\
K (NLTE)  	&  5.11	& ...	&  ---	&  -0.26 $^{*}$	&  0.27 $^{*}$	& 	0.09$^{*}$	&  0.40$^{*}$ & --- & 0.20	\\
Ca  	& 6.34	& 0.10	& 0.30	&-0.20	& 0.40	&	 0.30	& 0.30	& 0.00 & 0.30  \\
Sc 	& 3.15	& 0.25	& 0.00	& 0.70	& 0.40  &	 0.90	& 0.85	& 0.35 & 0.20 	\\
Rb  	& 2.52	& 0.60	&$<$-0.30	&$<$-0.50	& 0.70	& 	$<$ 0.10	& $<$0.10	& 0.60 & 0.20  	\\
Y   	& 2.21 	& 0.80	& 0.40	& 1.10	& 1.40	& 	 0.80	& 0.50	& 0.55 & 0.20 	\\
Zr  	& 2.58	& 1.30	& 0.40	& 1.50	& 1.20	& 	 1.50	& 1.70	& 0.80 & 0.20 	\\
Ba  	& 2.18	& 1.50	& 1.30	& 1.80	& 2.20	& 	 2.00	& 2.30	& 1.35 & 0.20 	\\
La  	& 1.10	& 1.70	& 1.40	& 2.00	& 2.20	&	 2.00	& 2.30	& 1.65  & 0.20  \\
Ce  	& 1.58	& 1.42	& 1.20	& 2.10	& 2.00	& 	 1.90	& 2.00	&1.07 & 0.20 	\\ 
Pr  	& 0.72	& 1.48	& 1.18	& 2.30	& 2.10	& 	 1.98	& 2.20	& 1.25 & 0.20 	\\
Nd  	& 1.42	& 1.48	& 1.20	& 1.90	& 2.10	& 	 1.66	& 2.20	& 1.35 & 0.20 	\\
Sm  	& 0.96	& 1.24	& 0.90	& 1.50	& 2.00	& 	 1.64	& 2.10	& 1.09 & 0.20 	\\
Eu  	& 0.52	& 1.06	& $<$0.36	& 1.16	& 1.60	& 	 1.66	& 1.96	& 1.23 & 0.20 	\\
$[hs/ls]$  & 0.00     &0.49   & 0.90   &0.70   &0.83   &    0.85   & 1.10   & 0.68         & ---\\
\hline
    \end{tabular}
    \label{tab:abund-chem}
    \begin{tablenotes}
        \item The * represents the values corrected for the NLTE effects and  --- represents no available NLTE correction.\\
        \item $^{\S}$ The Li abundance quoted  here is in the absolute scale, A(Li)
        \end{tablenotes}
\end{table*}
\begin{table}
\caption{The variation of abundances upon the variation of stellar parameters are mentioned here. The last column is the resulting change in the abundances due to the combined effect which is obtained by adding individual variations in quadrature.}
    \centering
    \begin{tabular}{ccccc}
    \hline
    \hline
    Element & \multicolumn{3}{c}{$\Delta log \epsilon(X)$} \\
    \cline{2-5}
    & $\Delta T_{eff}$ & $\Delta \logg$ & $\Delta\xi$ & $\sigma^{tot}_{sys}$ \\
    & ($\pm$150K )& ($\pm$0.25 dex) & ($\pm$0.2km s$^{-1}$ )& \\ 
 \hline
 \hline
    Li  &  $\pm$0.3  &  $\mp$ 0.15 &  $\pm$0.0 & 0.11 \\
O   &  $\mp$0.1  &  $\mp$ 0.05  & $\pm$0.0 & 0.01\\
Na  &  $\mp$0.05 &  $\pm$ 0.10  &  $\mp$0.05  & 0.02 \\
Al  &  $\mp$0.15 &  $\pm$0.10 & $\mp$0.05 & 0.04 \\
Mg  &  $\pm$0.0 & $\pm$0.1  & $\pm$0.0 & 0.01 \\
K   & $\pm$0.1 & $\pm$0.05  & $\pm$0.0  & 0.01 \\
Ca  & $\pm$0.1 & $\mp$0.1  & $\pm$0.0  & 0.02 \\
Sc  & $\pm$0.0 & $\mp$0.05  & $\mp$0.05 & 0.01 \\
Rb  & $\pm$0.1 & $\pm$0.0   & $\pm$0.05 & 0.01  \\
Y   & $\pm$0.05 & $\pm$0.05 & $\pm$0.0 &  0.01\\
Zr  & $\pm$0.1  & $\pm$0.0  & $\pm$0.0 &  0.01\\
Ba  & $\pm$0.15 & $\pm$0.1  & $\pm$0.0 & 0.03\\
La  & $\pm$0.0  & $\mp$0.05 & $\pm$0.0 & 0.01\\
Ce  & $\pm$0.05 & $\pm$0.05 & $\pm$0.0 & 0.01\\
Pr  & $\pm$0.15 & $\pm$0.05 & $\pm$0.05 & 0.03\\
Nd  & $\pm$0.05 & $\pm$0.05 & $\pm$0.0 & 0.01\\
Sm  & $\pm$0.1  & $\pm$0.1  & $\pm$0.0 & 0.02\\
Eu  & $\pm$0.2  & $\pm$0.1 & $\pm$0.0 &0.05 \\
\hline
\end{tabular}
    
    \label{tab:syserror}
\end{table}

\subsection{C, N and O}
\begin{figure}
    \centering
    \includegraphics[width = 0.48\textwidth]{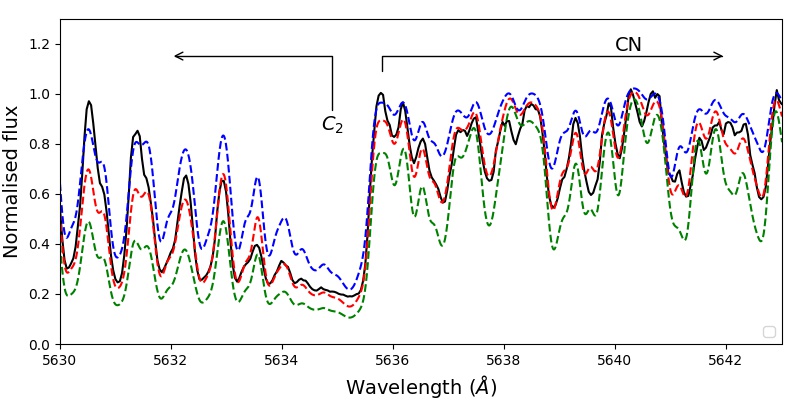}
    \caption{  The region of C$_{2}$ band head at 5635 \AA\ and the CN molecular features of HD 187216 are fitted with the synthetic spectra. The arrows indicate the region where the respective features are present. The  spectrum shown in black is the observed data, whereas the spectrum in red corresponds to the best fit value [C/Fe] = 1.45 and [N/Fe] = 0.93. Green and blue lines indicate syntheses corresponding to the range of uncertainty of $\pm$0.2  dex in both carbon and nitrogen abundances. }
    \label{fig:cn5635}
\end{figure}

\begin{figure}
\centering
\includegraphics[width = 0.48\textwidth]{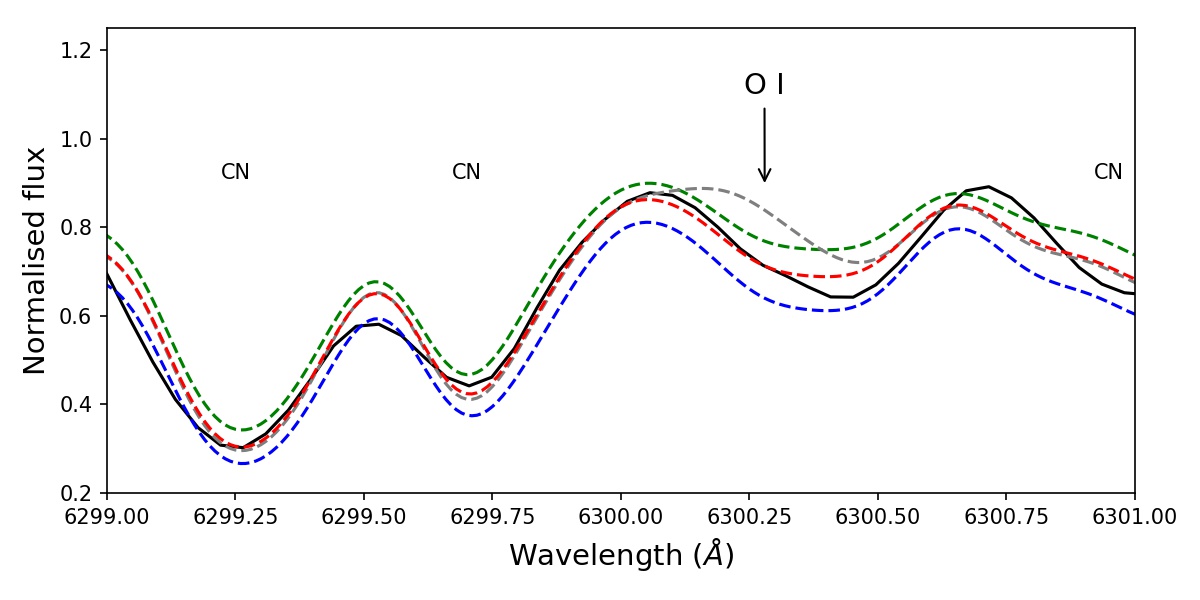}
\caption{The [OI] line  at 6300.3 \AA\ is fitted with synthetic spectra of different abundances. The black  solid line is the observed spectrum of HE 0017+0055 and the red spectrum is the best fit which corresponds to  [O/Fe] = 0.7 dex. The green and blue spectra denote the syntheses corresponding to the range of uncertainty of $\pm$ 0.3
dex in oxygen abundance.  The grey synthetic spectrum corresponds to the case where [O I] line has been removed from the line list.}
\label{fig:oxy6300}
\end{figure}

\begin{figure}
\centering
\includegraphics[width = 0.48\textwidth]{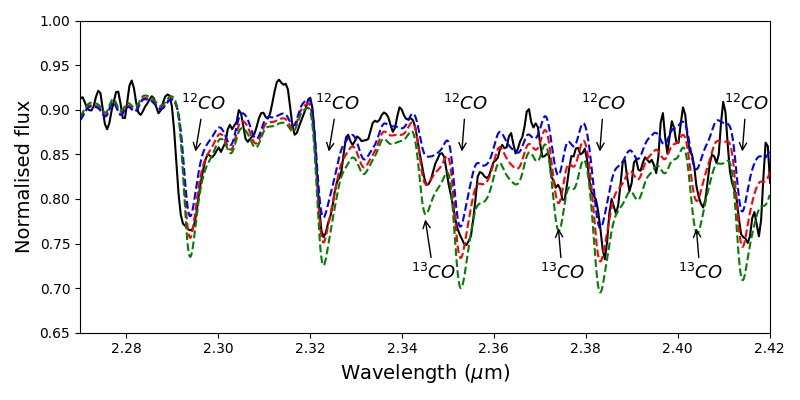}
\caption{The NIR CO molecular bands are fitted with various synthetic spectra. The black solid line represent the observed spectrum of HE 0017+0055. The red spectrum is the best fit to the observed spectrum which corresponds to [C/Fe] = 2.1 and [O/Fe] = 0.6. The green and blue spectra indicate syntheses corresponding to the range of uncertainty of $\pm$ 0.3
dex in oxygen abundance. }
\label{fig:nirco}
\end{figure}

\begin{figure}
    \centering
    \includegraphics[width = 0.48\textwidth]{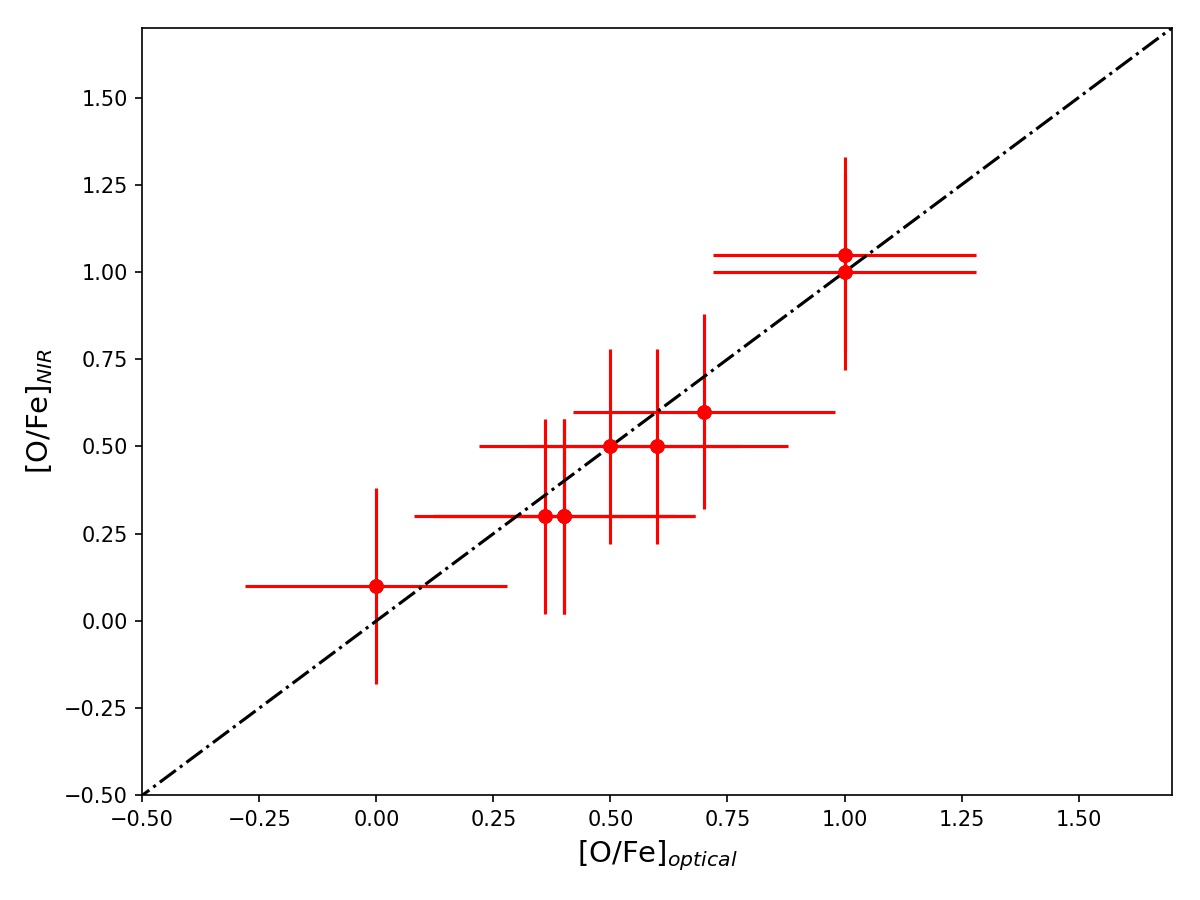}
    \caption{The oxygen abundances derived from optical and NIR spectra have been compared. The abundances are matching within an error of 0.3 dex. The black dotted dashed line corresponds to the points where the oxygen abundances from NIR and optical spectral lines are identical.}
    \label{fig:comp_oxy}
\end{figure}

The carbon abundance in the program stars has been derived by fitting the
C$_{2}$ molecular band heads at 5165 \AA\ and 5635 \AA. Both
features yielded the same abundance values and we adopted this carbon value while deriving the abundances of N and O from molecular features.
 For the case of N abundance,  we used CN lines in the wavelength
range from 5635-6700 \AA\ by iteratively changing the nitrogen
abundance of the synthetic spectra by 0.2 dex and fixed the carbon abundance to  the value derived from C$_{2}$ molecular band and fitting the entire spectral region (refer figure \ref{fig:cn5635}). The best-fit value is chosen as the final nitrogen abundance. We did not use strong CN lines in the blue region at 4215 \AA\, due to the poor the SNR .  \\
The oxygen abundance from optical spectra  was derived by using the [O I] lines at 6300, 6363 \AA\ . Since these lines are
blended with CN features, a small change in the nitrogen abundance affects the derived oxygen abundance. So, the carbon
abundance and the nitrogen abundance were first fixed to be the value obtained from
the C$_2$  and CN bands and the oxygen abundance was iteratively adjusted to fit the [O I] lines. While fixing the O abundance, N abundance was also iteratively changed by a small amount for a better fitting of the [OI] feature but without affecting the nearby CN features. The resulting O abundance from both the lines
differs by 0.3 dex because of the iterative process involving both
N and O, and we take this difference as our final uncertainty on
the reported oxygen abundances. The spectral fitting of the [OI] line is shown in the figure \ref{fig:oxy6300}.
The oxygen abundance from NIR CO molecular features at 2.3 $\mu$m were also obtained for all the program stars in this study. The C and N abundance derived from optical spectra were kept fixed in the synthetic spectral grid and we varied the oxygen abundances with a step size of 0.2 dex and the best fit value was chosen as the final value (see figure \ref{fig:nirco}). 
The final C, N and O abundances were listed in the table $\ref{tab:abund-chem}$ and the oxygen abundances derived from both optical and NIR spectra are compared in the figure \ref{fig:comp_oxy}. The O abundance from optical and NIR are matching within an uncertainty of 0.3 dex. So the average of these two values was used as O abundance elsewhere in the text. 
 
\subsection{Li and $^{12}C/^{13}C$ ratio}
\begin{figure}
\centering
\includegraphics[width= 0.48\textwidth]{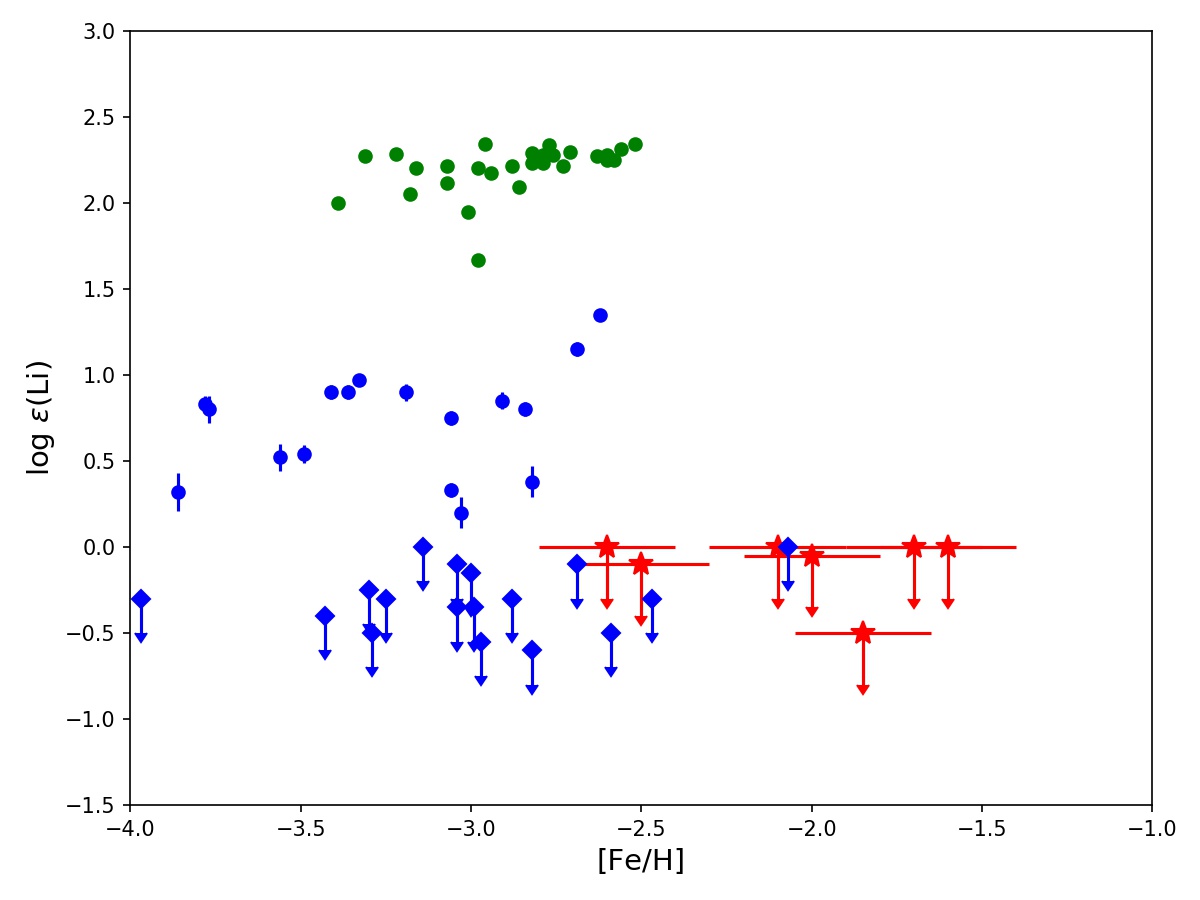}
\caption{The Li abundance of our program stars (red star symbols) is compared with the Li abundance of C-normal stars from the literature. The filled circles correspond to unmixed stars from \citet{Sbordone2010} (green, main-sequence turnoff and dwarf stars) and \citet{spite2005} (blue, early RGB) whereas the blue diamond symbols correspond to the mixed giants from \citet{spite2005}. The Li abundance of our program stars occupy  the same region as that of the mixed giants.} 
\label{fig:licompare}
\end{figure}

\begin{figure}
\centering
\includegraphics[width= 0.48\textwidth]{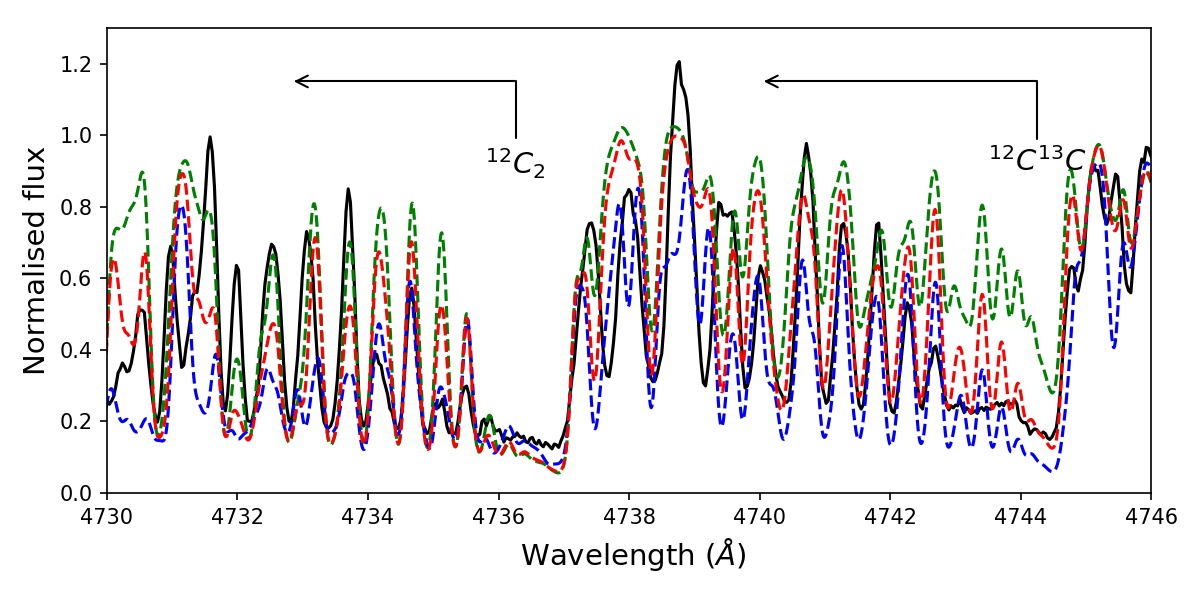}
\caption{The $^{12}C_{2}$ and $^{12}C^{13}C$ molecular features of HE 0017+0055 are plotted as black spectrum. The bandheads at 4737 and 4744 \AA\ are marked with arrows indicating the direction at which the corresponding molecular features are present. From the figure, it can be seen that the red spectrum is the best fit and it corresponds to a $^{12}C/^{13}C$ = 9. The blue spectrum corresponds to $^{12}C/^{13}C$ = 2 whereas the green spectrum corresponds to $^{12}C/^{13}C$ = 99.} 
\label{fig:c2c3optical}
\end{figure}

\begin{figure}
\centering
\includegraphics[width = 0.48\textwidth]{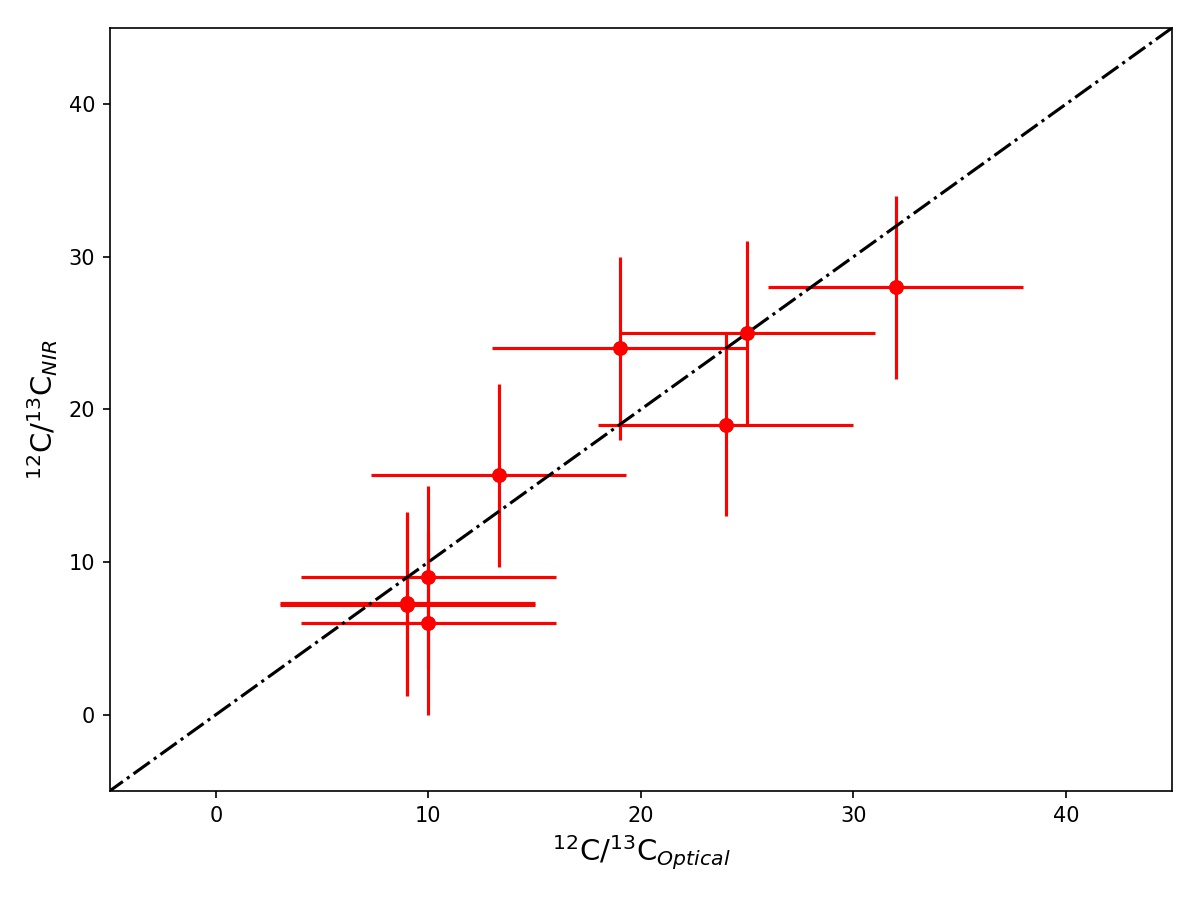}
\caption{The carbon isotopic ratio derived from optical and NIR spectra are compared here. The black dotted dashed line corresponds to the points where the $^{12}C/^{13}C$ from NIR and optical spectral lines are identical.}
\label{fig:c1213}
\end{figure}

 Li-abundances for the program stars were determined by synthesizing
the resonance doublet at 6707 \AA\ and the abundances are provided in the table \ref{tab:abund-chem}. We did not detect any distinguishable spectral features of Li in the region, thus, we only derived an upper limit for the abundances. Due
to the heavy contamination from CN molecular features to the Li line,  we assigned an uncertainty of
0.3 dex to the derived values. Since all the samples in the study are in the giant phase, the low values for A(Li) \footnote{A(X) = log $\epsilon$(X) = log (N$_{X}$/N$_{H}$) + 12, where N$_{X}$ and N$_{H}$ represent
number densities of a given element X and hydrogen  respectively.} are consistent with their evolutionary phase.  During the RGB evolutionary stage, the surface Li is taken into the deeper interiors through convective mixing where it is easily destroyed due to the high interior temperatures \citep{gratton2000aa, lind2009}.  To confirm the mixing scenario, we compared the Li abundances of our program stars with the objects from \citet{spite2005} and \citet{Sbordone2010} in the figure \ref{fig:licompare} and identified that all the program stars occupy the same region as that of the mixed giants.\\
The $^{12}C/^{13}C$ ratio was calculated using the NIR $^{12}CO$  and $^{13}CO$ molecular band heads at 2.38 $\mu$m (refer figure \ref{fig:nirco}). We have also derived the carbon isotopic ratio from $^{12}C_{2}$ and $^{12}C^{13}C$ band heads at 4737 and 4744 \AA\ respectively (refer figure \ref{fig:c2c3optical}) to check the consistency of the values derived from low-resolution NIR spectra. We assign an uncertainty of 6 to the  $^{12}C/^{13}C$ ratio measurements, below which the variation in the ratio could not be detected with the goodness of fit. 
When comparing, we found that the $^{12}C/^{13}C$ ratios from optical spectra are matching with that derived from the NIR $^{12}CO$  and $^{13}CO$ molecular bands and compared these in the figure \ref{fig:c1213}. The ratios derived from optical and NIR spectra are mentioned in table \ref{tab:comparison}.
\subsection{ $\alpha$-elements and odd Z elements }
Among the $\alpha$-elements, we derived the abundances of Mg and Ca. The Mg I  triplet lines at 5172 \AA\ and 5183 \AA\  lines are very strong so we did not use them to derive the Mg abundance. Instead, we have used the 5528 \AA\ and 5711 \AA\ to derive the abundances, in particular, we have used 5711 \AA\ for the abundance measurements only if the line is identified to be free from blending.  The NLTE corrections for these lines are either negligible or not available for the parameter set of our program stars \citep{Lind2011, bergemann2017a, bergemann2017ApJ16B}.  We have used four Ca I lines, whichever are free of blending with molecular features, to derive the Ca abundance. The median of the abundances from individual lines is calculated and provided in the table \ref{tab:abund-chem}. \\
Among the odd Z elements, we derived the abundances of Na, Al, K and Sc using the high resolution spectra and the details of the lines are given in the table \ref{tab:linelist}. The Na D lines were not used as the lines are too strong for deriving the abundances. So we depended on the weaker lines 5682 \AA\ and 5688 \AA\ to derive the Na abundance. The NLTE corrections for these Na I lines were performed from \citet{Lind2011} whenever available and the average  values are listed in the table \ref{tab:abund-chem}. Al I lines at 6696 and 6698 are used for deriving the Al abundance and these lines have negligible contributions from NLTE effects \citep{Baumueller1997,nordlander2017aa}. For the case of K, whenever the feature  at 7698.98 \AA\ has contributions from telluric features, the abundance from 7664.87 \AA\ is quoted,  otherwise the average of the abundances from both the lines is quoted in table \ref{tab:abund-chem}. Both these lines are sensitive to NLTE effects \citep{takeda2002PASJ,Kobayashi2006ApJ, Andrievsky2010AA, prantzos2018MNRAS, reggiani2019aa} and the NLTE corrections depend on the effective temperature and surface gravity of the model. Using the NLTE grid provided in \citet[]{reggiani2019aa}, we have estimated the NLTE corrections for the program stars and the NLTE corrected values are given in the table \ref{tab:abund-chem}. The NLTE grid does not cover very low \logg\ values, so we could not estimate the NLTE corrections for HE 1152-0355 and HE 0314-0143.  We could not measure the K abundance in HD 5223 due to the contamination from telluric features. For the case of Sc abundance, we  could not find any study of Sc line  formation in NLTE except in \citet{Zhang2008aa}  where they have studied the NLTE effects only for the sun and identified negligible NLTE effects for Sc II lines. So we report only LTE results for this element in the table \ref{tab:abund-chem}.  The resolution of the  NIR spectra was too low to resolve  the HF feature from the nearby features, so we could not derive F abundance in these stars.
\\
\\
The spectral features of the Fe-peak elements are  severely blended with the molecular features of C and N to derive any meaningful abundances. So we could not make any reliable measurements of their abundances.
\subsection{Neutron capture elements}
We  derived the abundances of light $\it {s}$-process elements (ls: Y, Zr)  as well as heavy $\it{s}$-process elements (hs: Ba, La, Nd, Ce, Pr, ) and r-process elements(Eu and Sm) to understand the origin of neutron-capture elements using various spectral lines in the high-resolution optical spectrum. The [hs/ls] \footnote{ This ratio defines the enrichment of the second peak of the $\it{s}$-process (hs),  with respect to the first peak, light $\it {s}$-process (ls) elements.  It is defined
as [hs/ls] = [hs/Fe]- [ls/Fe] in the usual spectroscopic notation.We adopt [ls/Fe] = [Y+Zr/Fe]/2 and [hs/Fe] = [Ba+La+Ce/Fe]/3.} ratio of all the program stars are provided in the table \ref{tab:abund-chem}.  
The CN band head at 4215 \AA\ is saturated in all our program stars. So the abundance of Sr could not be measured from the Sr II resonance line at 4216 \AA. The details of the lines used for the abundance measurements are given in the table \ref{tab:linelist}. 
For the case of Rb abundance, we have quoted the abundance from the Rb I line at 7800.2 \AA\ and we did not use the 7947.5 \AA\ line since it was severely blended with the CN molecular feature. For Ba and Eu, we used the Ba II lines at  5853 \AA\ and 6141 \AA\ and Eu II line at 6645 \AA\ for the abundance determination. We checked for available NLTE corrections to these features and identified to have negligible contributions from NLTE effects  \citep{mashonkina1999aa, mashonkina2000aa, Mashonkina2014, gallagher2020}.
\subsection{Oxygen abundance of the program stars }
We derived the oxygen abundance of 9 carbon enhanced stars using both optical and NIR spectra.  The oxygen abundances of  HD 5223, HD 187216, BD+42 2173 and C*782 are not available so far in the literature.  For the case of HE 1152-0355, HE 1418+0150, HE 0017+0055 and HE 0314-0143; \citet{kennedy2011aj} had reported the oxygen abundances from the low-resolution NIR CO lines. Using low-resolution NIR and high resolution optical observations, we confirm the O abundances in these stars. The abundances are given in table \ref{tab:comparison}. The stellar parameters and  abundance details of C*782 and BD+41 2150 are not discussed elsewhere in the paper as their abundance pattern was found to be more complex and does not fit in the theme of this paper.
So we confine the results from these two stars to the oxygen abundance and their carbon isotopic ratio. A detailed abundance analysis of the two objects will be discussed in an upcoming paper, Susmitha {\it et al.} (in prep).
\begin{table}
\centering
\caption{The oxygen and carbon isotopic ratio from NIR and optical spectra.}
\label{tab:comparison}
\begin{tabular}{|c|c|c|c|c|}
\hline
star name& \multicolumn{2}{c}{[O/Fe]} & \multicolumn{2}{c}{$^{12}C/^{13}C$}\\
\cline{2-5}
        & Optical & NIR & Optical & NIR \\
\hline
\hline
 HD 5223        & 0.4   & 0.3 & 25  & 25    \\
HE 1152-0355    & 0.5   & 0.5 & 10  & 9     \\
HE 1418+0150    & 1.0   & 1.0 & 9   & 7   \\
BD+42 2173      & 0.4   & 0.3 & 39  & 32    \\
HD 187216       & 0.0   & 0.1 & 10  & 6     \\
HE 0017+0055    & 0.7   & 0.6 & 9   & 7    \\
HE 0314-0143    & 1.0   &1.1 & 13  & 16    \\
BD+41 2150  &  0.6 & 0.5 & 24 & 19 \\
C* 782         & 0.4 & 0.3  & 32 & 28  \\
\hline
\end{tabular}
\end{table}
\subsection{Comparison with literature}
Since our program stars were chosen from the literature, we compare the stellar parameters and various elemental abundances with the previous studies . \\
{\bf HD5223:} This program star was studied by \citet{Goswami2006mnras} and classified as a \cemps\ star based on the abundances from Ba, La and Ce. The derived stellar parameters from their study and from our study are matching within the error quoted in sec. \ref{error}. But the classification of HD 5223 as \cemps\ stars is revisited here and classified it as \cemprs\ using the [Eu/Fe] abundances which was not done in the earlier studies. \\
{\bf HE 1152-0355:} \citet{Goswami2006mnras} have reported the stellar parameters and classified the star as a \cemps\ star. \citet{kennedy2011aj} have re-derived the stellar parameters and obtained the C and O abundances. Our stellar parameters and C and O abundances are also matching with the values from  \citet{kennedy2011aj} within the uncertainties quoted in table \ref{tab:abund-chem}. \\
{\bf HE 1418+0150:} \citet{kennedy2011aj} have reported the stellar parameters and C and O abundances using low-resolution spectra. 
We derived the stellar parameters and abundances using high resolution spectra and obtained a larger \logg\ and slightly higher metallicity (refer table \ref{tab:stparam}).  \citet{cjhansen2016} has derived the Sr and Ba abundances of this star and classified them as MP({\it s}). The abundances of rest of the neutron-capture elements in this star and their nucleosynthetic origin were not reported earlier.  We classify the star as \cemprs\ using the [Eu/Fe] and [Ba/Fe] values.\\
{\bf HD 187216:} \citet{kipper1994aa} reported this star as a  possible intrinsic AGB star due to the high carbon and $\it {s}$-process element abundances with a low \logg\ value and the non-detection of a  binary companion. \citet{jorissen2016aaBinaryCHstars} monitored the radial velocity and confirmed the binary nature with a period longer than 10$^4$ days. We classify this stars as \cemprs\ star based on the abundances of Ba and Eu.
\\
{\bf HE 0017+0055:} This star was identified as a \cemprs\ star using the excess abundance of C and $\it {s}$-process and r-process elements and radial velocity variations with a period of 384 days \citep{jorissen2016aaBinaryCHstars, jorissen2016aa}. \citet{kennedy2011aj} derived the O abundance in the star using NIR CO lines. Our stellar parameters and the various elemental abundances are also inline with the estimates from   \citet{kennedy2011aj} and \citet{jorissen2016aa}. 
\\
{\bf HE 0314-0143:} \citet{kennedy2011aj} derived the stellar parameters, C and O abundances from low-resolution spectra. In the literature, this star was not classified into any sub category of carbon enhanced stars as no high-resolution observations were available. We classify this star as \cemprs\ based on the neutron-capture elemental abundances. 
\\
{\bf BD+42 2173:} \citet{mcclure1990apj} identified this star showing radial velocity variations with a period of 328 days. \citet{aoki1997aa} reported lower $^{12}C/^{13}C$ value whereas we have obtained a slightly higher value for $^{12}C/^{13}C$ ratio which is given in table \ref{tab:comparison}. 
\section{Discussion}
\begin{figure}
\centering
\includegraphics[width = 0.48\textwidth]{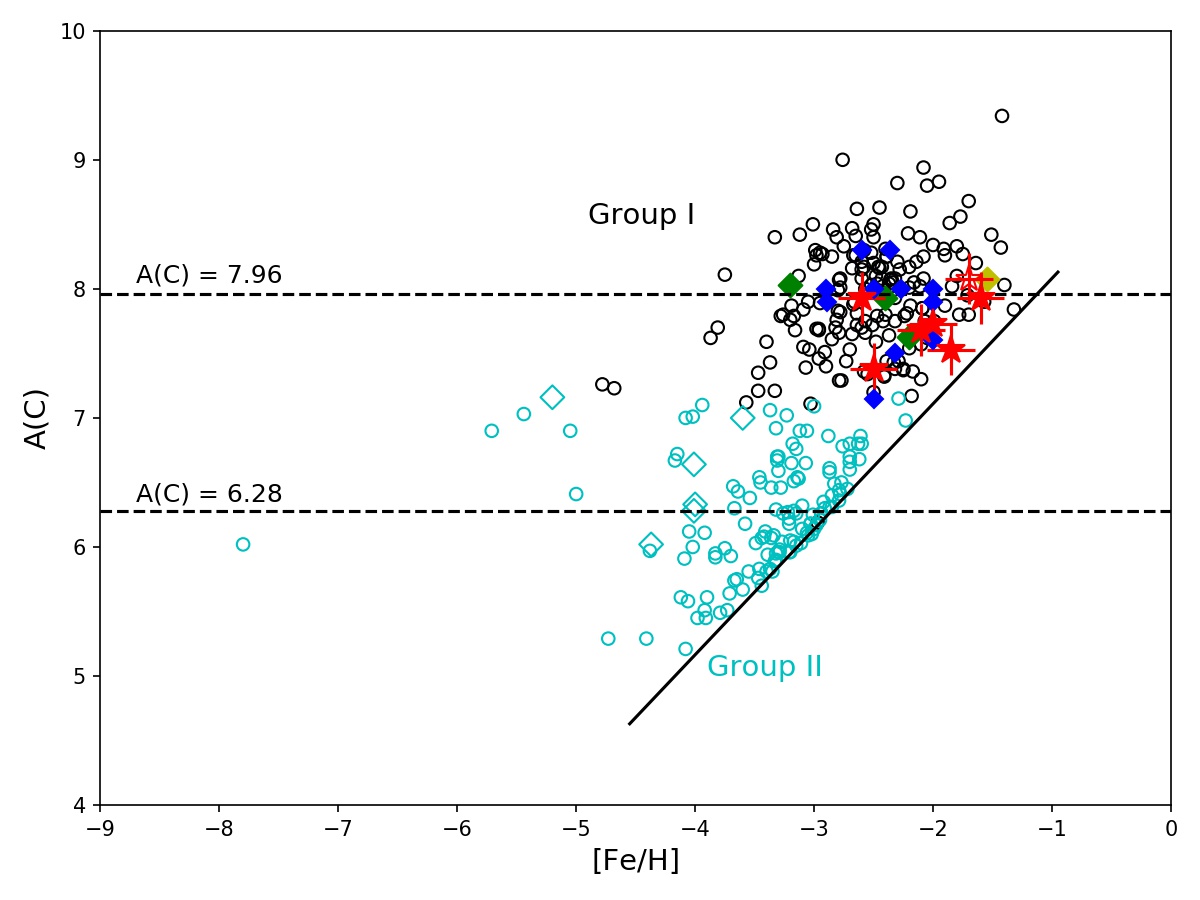}
\caption{ The C abundance of CEMP stars from  \citet[\& references therein]{Yoon2016} is plotted as open circles in which the CEMP${-s}$ and ${-r/s}$ stars are plotted as black open circles  whereas the CEMP-$no$ stars are plotted as cyan open circles. The filled  symbols represent the C abundance of \cemprs\ stars from various studies including \citet[green diamonds]{hansencj2019aa}, \citet[yellow diamond]{sbordone2020aa} and \citet[blue diamonds]{drisya2021aa}. The red star symbols correspond to the C abundances of stars from this study. The cyan open diamond symbols represent the C abundance of stars from \citet{bonifacio2018aa}.  The black solid line corresponds to [C/Fe] = +0.7. The horizontal dashed lines represent the high and low C band according to \citet{Yoon2016}. The \cemprs\ stars are identified to be scattered in the high C band region.}
\label{fig:yoonplot}
\end{figure}

We studied the abundance pattern of 7 CEMP stars using optical and NIR spectroscopy. These stars  exhibit high carbon abundance.   \citet{spite2013aa, bonifacio2015aa, hansentt2015apj}, and \citet{ Yoon2016} identified the existence of bimodality in  the distribution of A(C) for CEMP stars where most of the \cemps\ stars populate the high-C band whereas most of the \cempno\ stars populate the low-C band.  We compared the C abundance of our program stars with the CEMP stars  from the literature \citep[\& references therein]{Yoon2016,bonifacio2018aa, hansencj2019aa,sbordone2020aa,drisya2021aa} to check any peculiar abundance trend is associated with \cemprs\ subcategory (refer figure \ref{fig:yoonplot}). From the figure, it can be seen that the C abundance of \cemprs\ stars populate the high-C band region in the A(C) {\it vs} [Fe/H] space. While the  A(C)-[Fe/H] diagram is useful in separating the \cempno\ stars from other CEMP subclasses, it is not efficient in separating  \cemprs\ stars from \cemps\ sub-category. 
The program stars also exhibit enhancement in the neutron-capture elements. The radial velocities measured from the current optical spectra and the literature values, show variations indicating the possibility of the program stars to be in a binary system. The enhancement in carbon along with the neutron-capture elements and radial velocity variations favor AGB mass transfer star as a possible mechanism for the enhanced carbon and n-capture elements. The extrinsic origin of C is  further supported by the location of A(C) of these stars in the A(C)- [Fe/H] diagram. 
The abundance pattern exhibited by the CEMP stars provide clues about AGB nucleosynthesis at such low-metallicities. 
Here, we use the n-capture elements to understand the AGB nucleosynthesis and various mixing processes that will bring the carbon and neutron rich nuclei to the surface, while the abundance of light elements such as C,N,O and Na can be used to constrain the mass of the companion.
So we divide the discussion into two parts. In the first part, we will show how the abundance of light elements such as C, N, O and Na can be used to constrain the mass of the companion AGB star and in the second part, we will discuss how the neutron capture elements can provide details of the inter-shell mixing and nucleosynthesis processes.
\subsection{C,N,O and Na to constrain the AGB mass}\label{cnoagbmass}
\begin{figure*}
\centering
\begin{tabular}{cc}
\includegraphics[height = 5cm, width= 7cm]{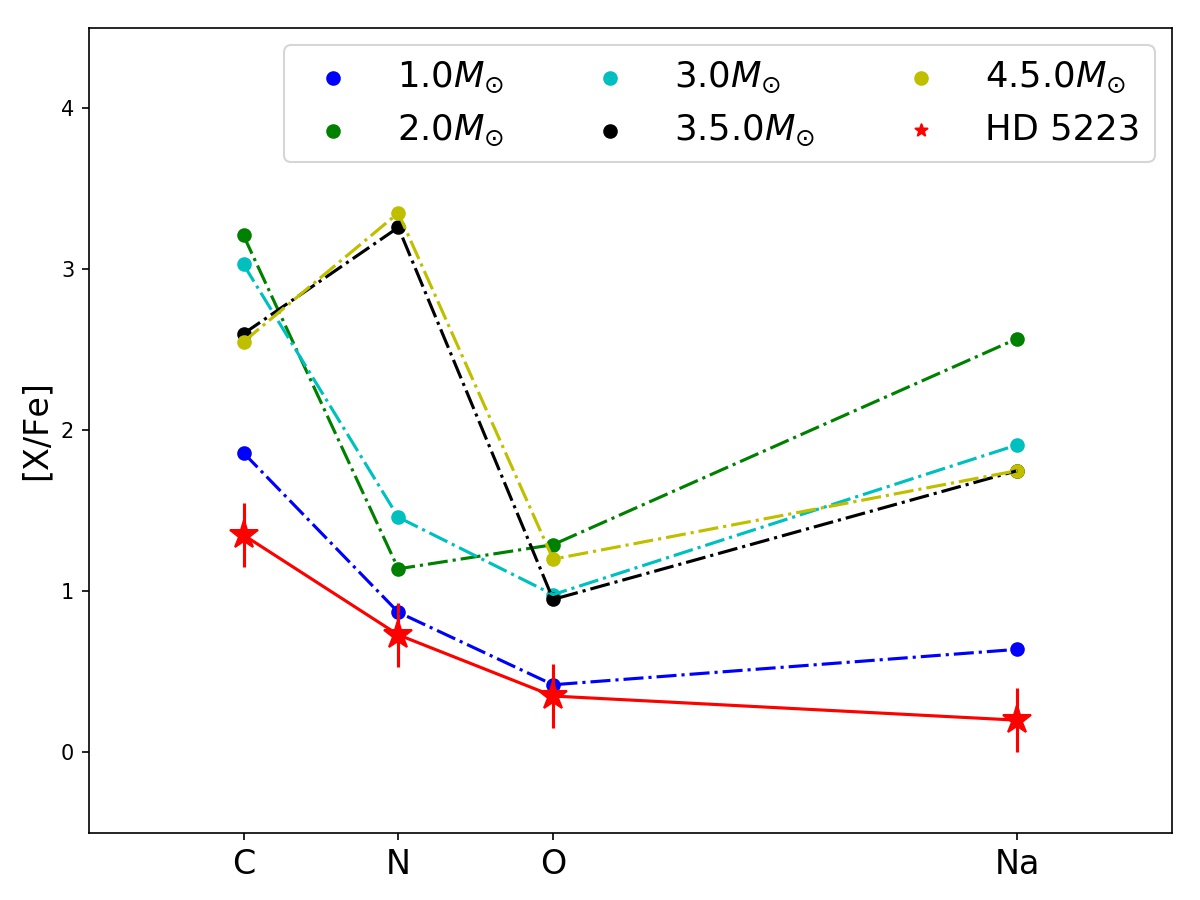}
     &  \includegraphics[height = 5cm, width= 7cm]{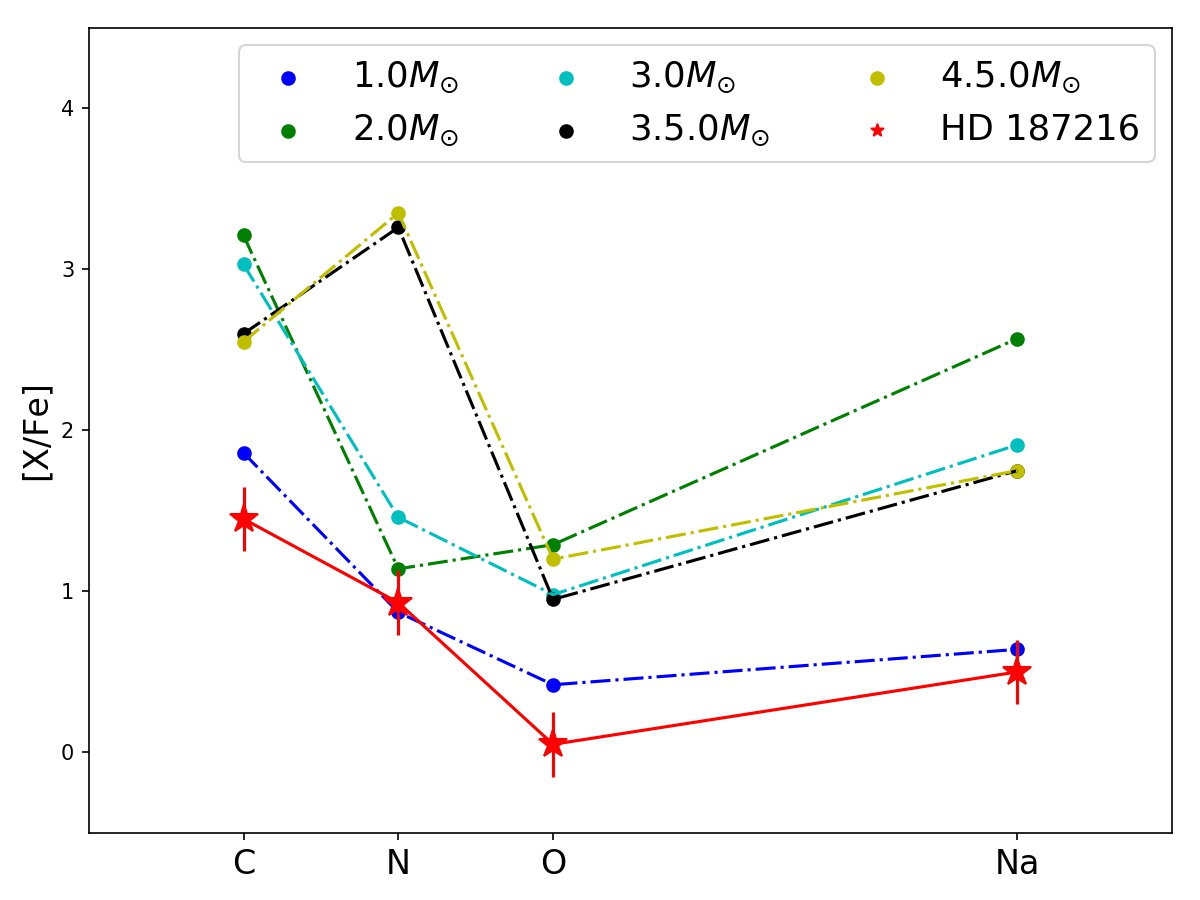} \\
    \includegraphics[height = 5cm, width= 7cm]{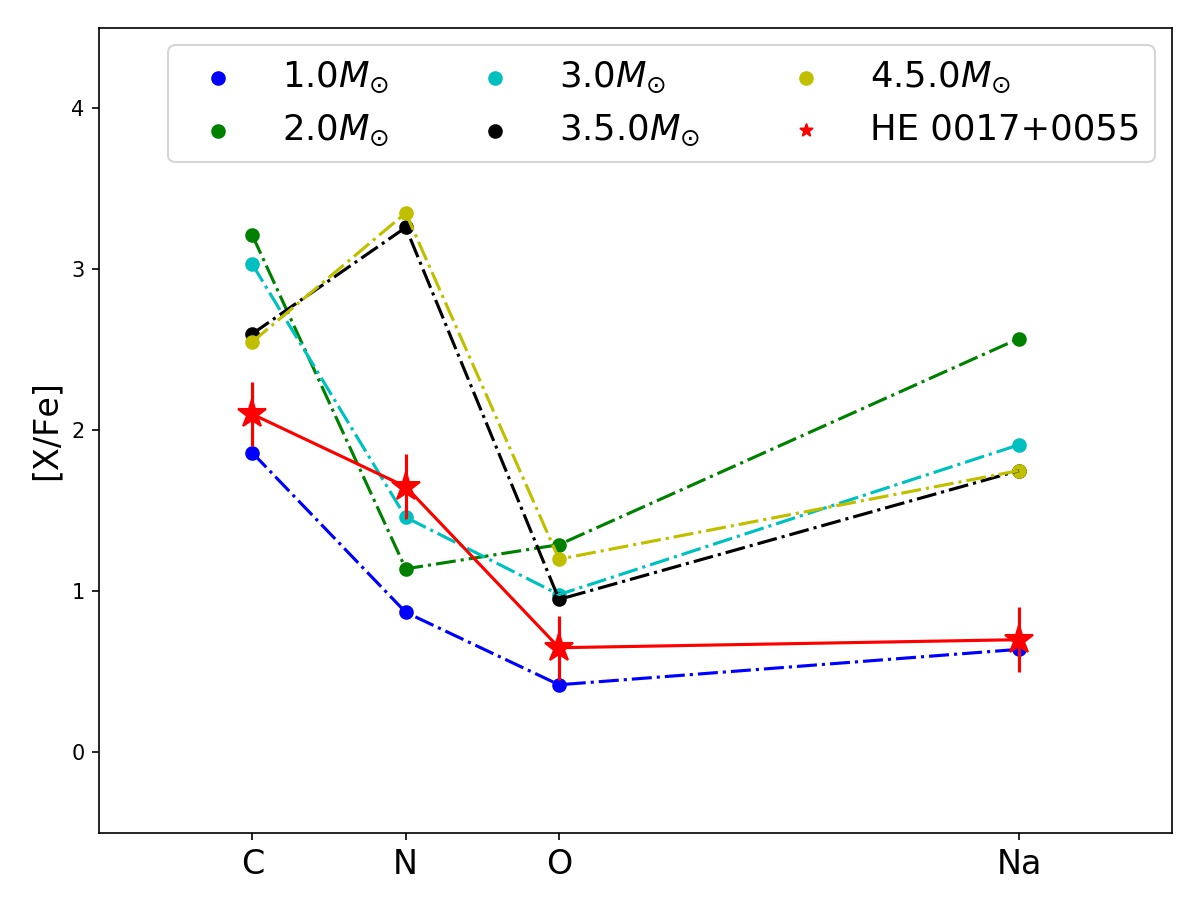} &
    \includegraphics[height = 5cm, width= 7cm]{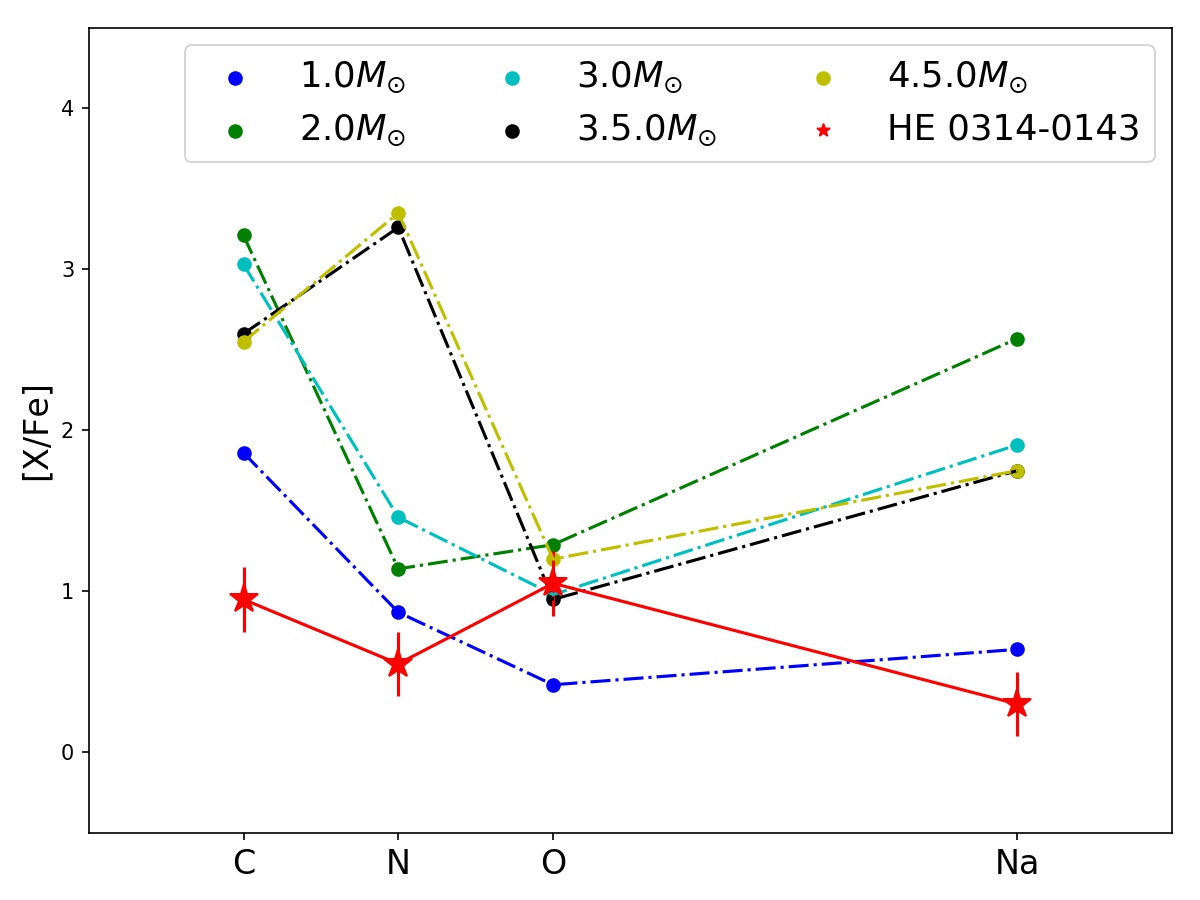} \\
    \includegraphics[height = 5cm, width= 7cm]{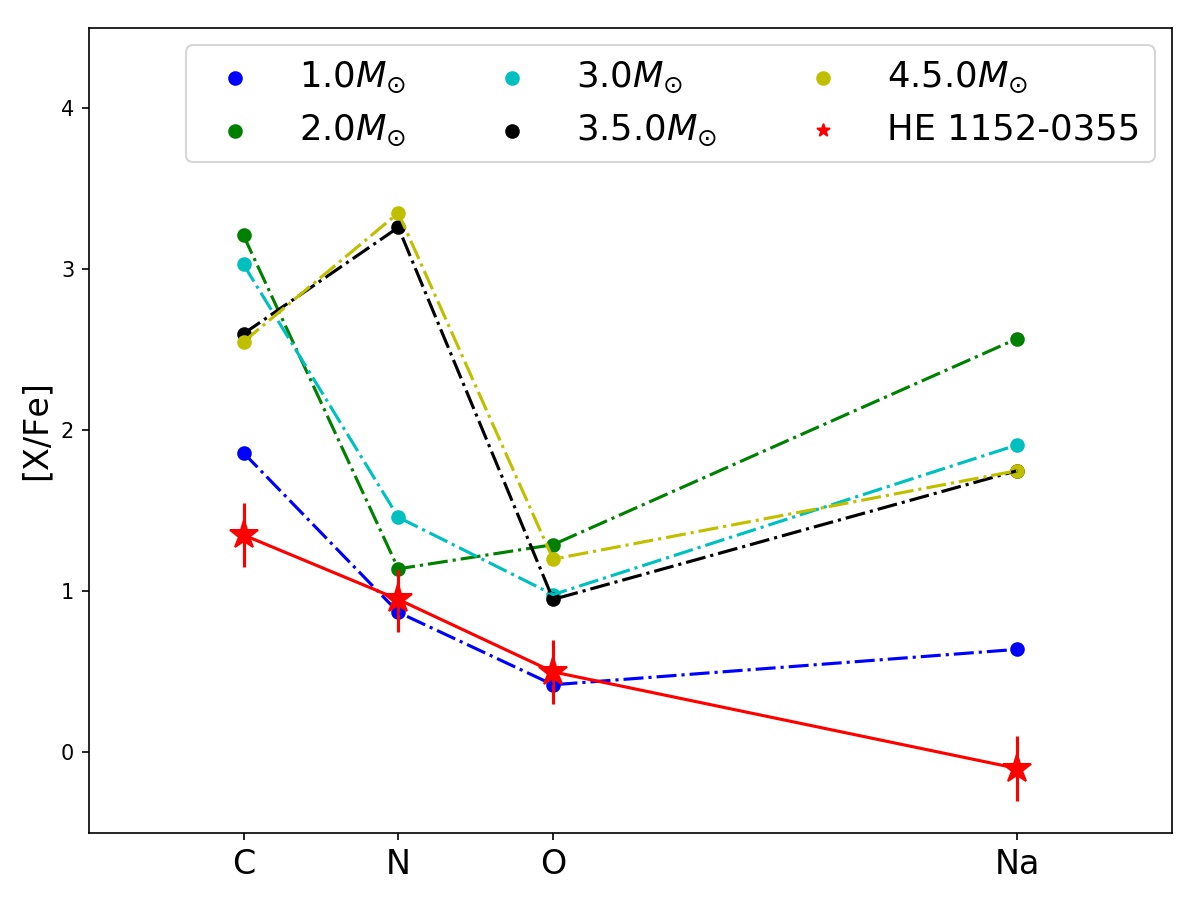} &
    \includegraphics[height = 5cm, width= 7cm]{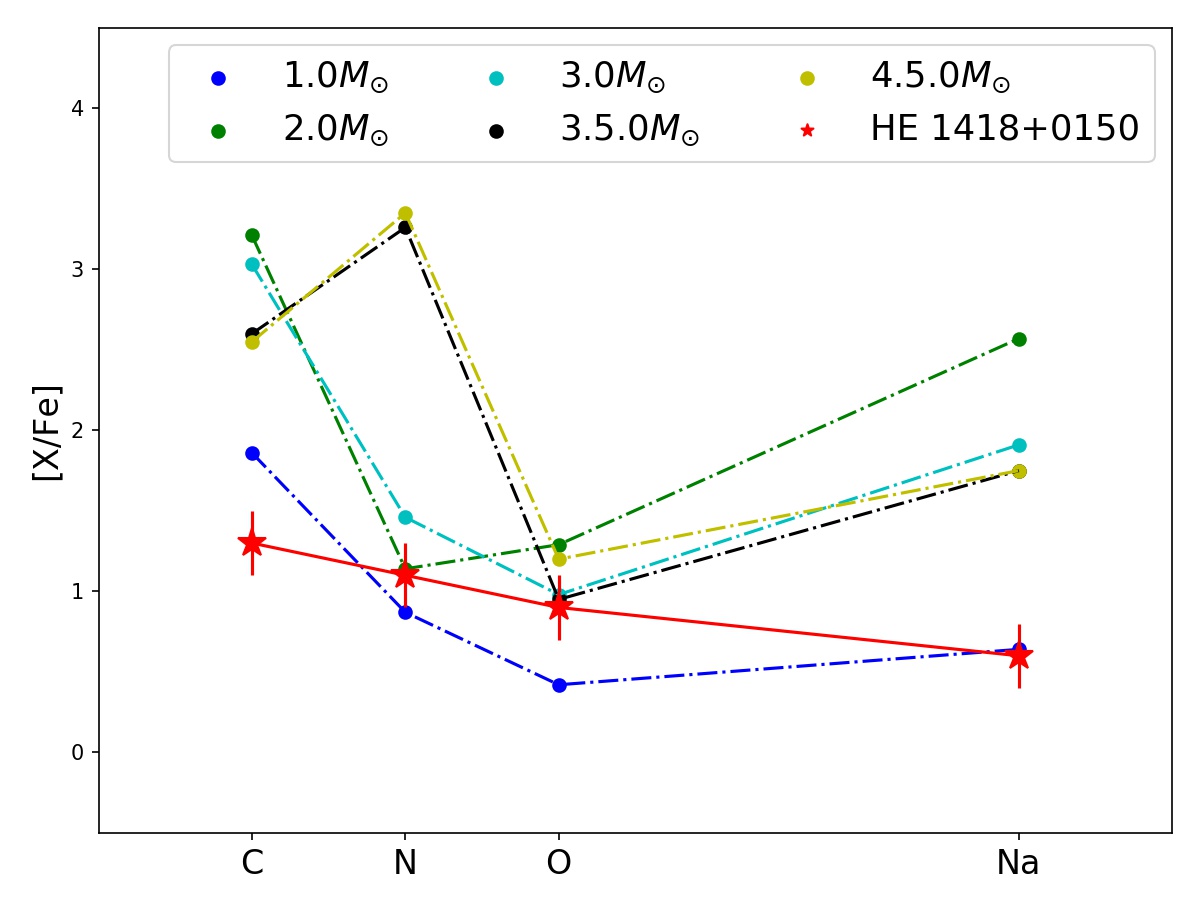} \\
    \includegraphics[height = 5cm, width= 7cm]{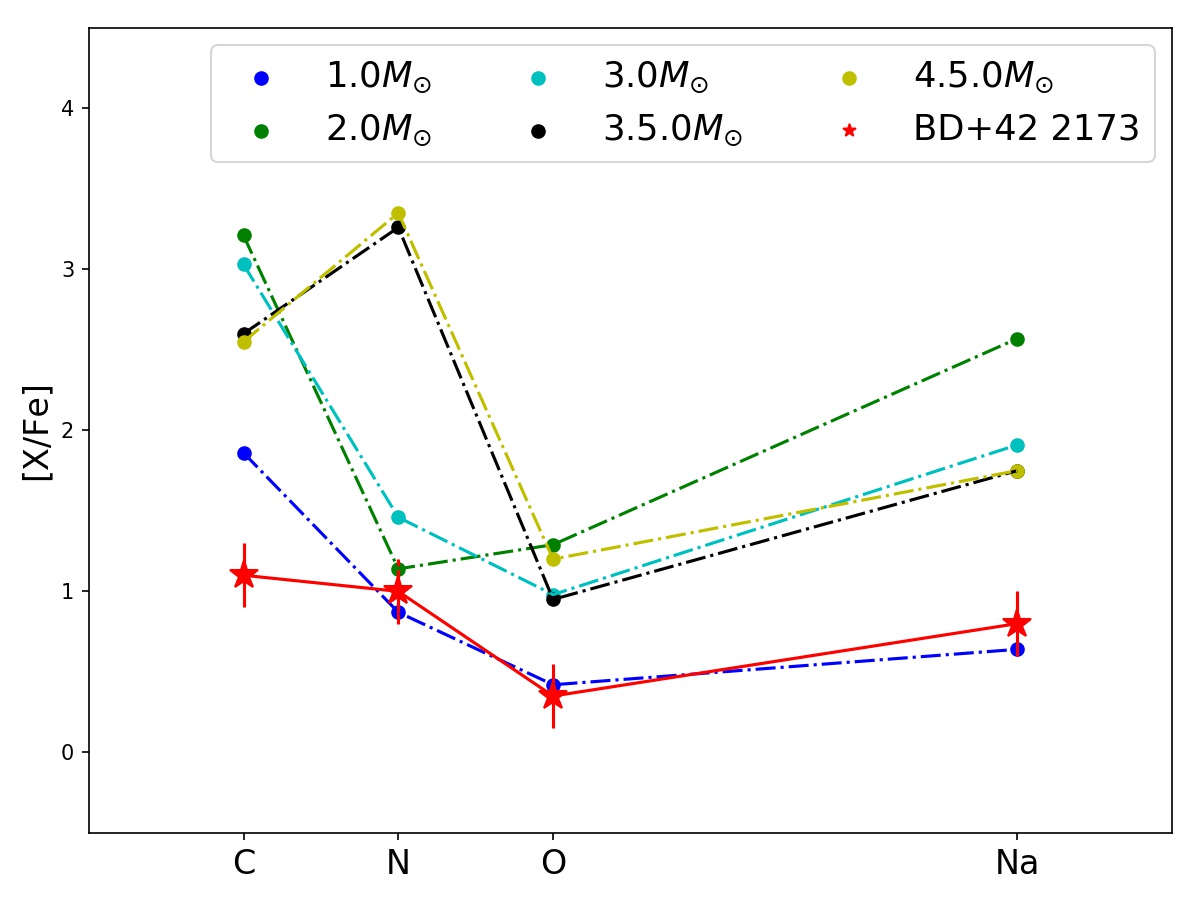} & \\
\end{tabular}
\caption{ The observed CNO and  Na(LTE) abundances of the program stars and the abundances from the standard AGB models
of different masses \citep{lugaro2012apj} are plotted. The observed values are not corrected for the dilution and mixing of CN processed material in the CEMP star. However,
it can be seen that the trend of the abundance distribution of the observed stars
resembles with that of low-mass  AGB stars (M $\leqslant$ 3M$_{\odot}$). For the case of HE 0314-0143,( which has the lowest temperature,  lowest \logg\  among the program stars), the poor match between the observed values and models could be due to some extra mixing during the evolution  which would result in a larger depletion of C.}
\label{fig:cno-comp-agb}
\end{figure*}

\begin{figure}
\centering
\includegraphics[width = 9.0cm, height= 7cm]{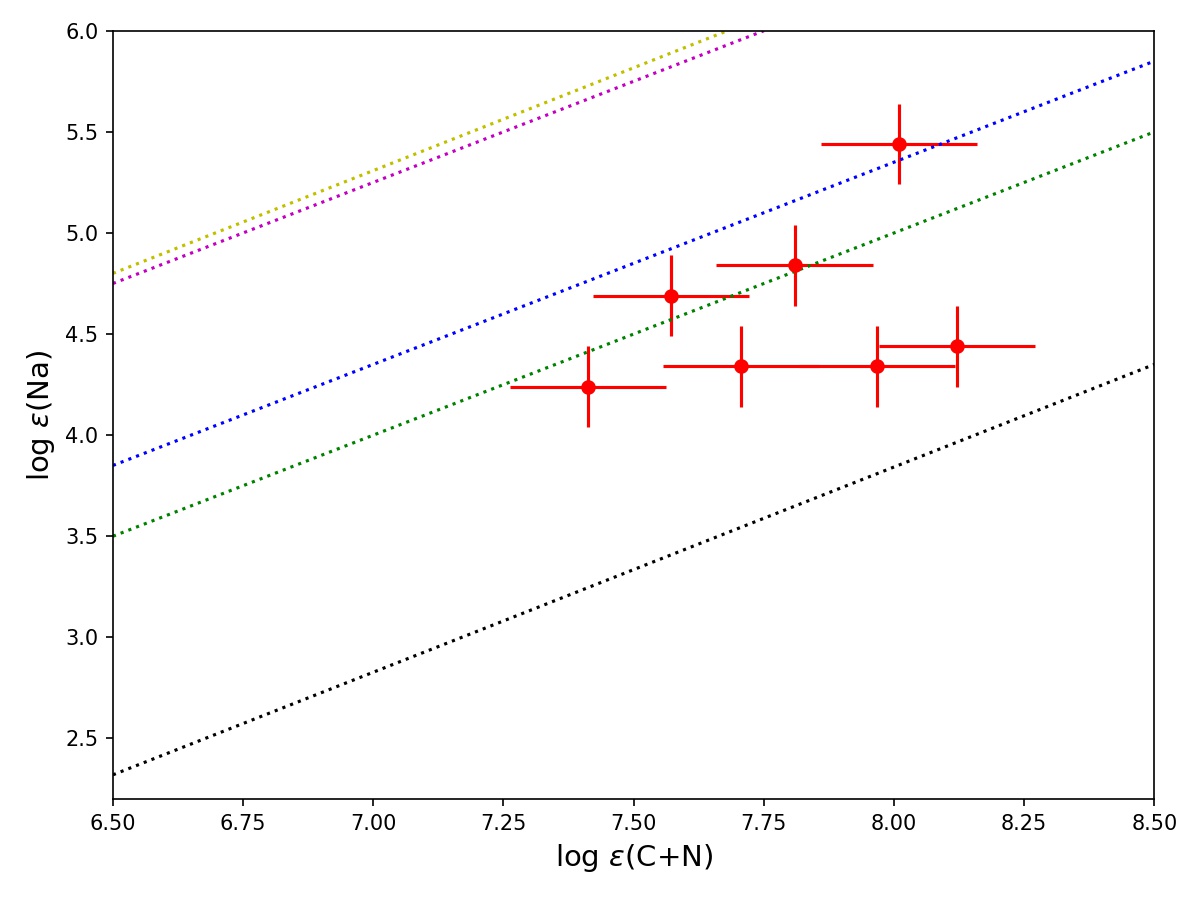}
\caption{The absolute Na abundances (LTE) of our program stars are  plotted as a function of the absolute (C+N) abundances. The dotted lines represent the abundances corresponding to AGB models from \citet{karakas2007pasa} for different masses: the black dotted line corresponds to the AGB model of 1.25M$_{\odot}$, the green dotted line corresponds to 1.75M$_{\odot}$,the blue dotted line corresponds to  2.5M$_{\odot}$, the magenta dotted line corresponds to 3.0M$_{\odot}$ and  the yellow dotted line corresponds to 3.5M$_{\odot}$ AGB model.  This figure is inspired from figure 10 in \citet{lucatello2011apj}.}
\label{fig:nacplusn}
\end{figure}

\begin{figure}
\centering
\includegraphics[width = 9.0cm, height= 7cm]{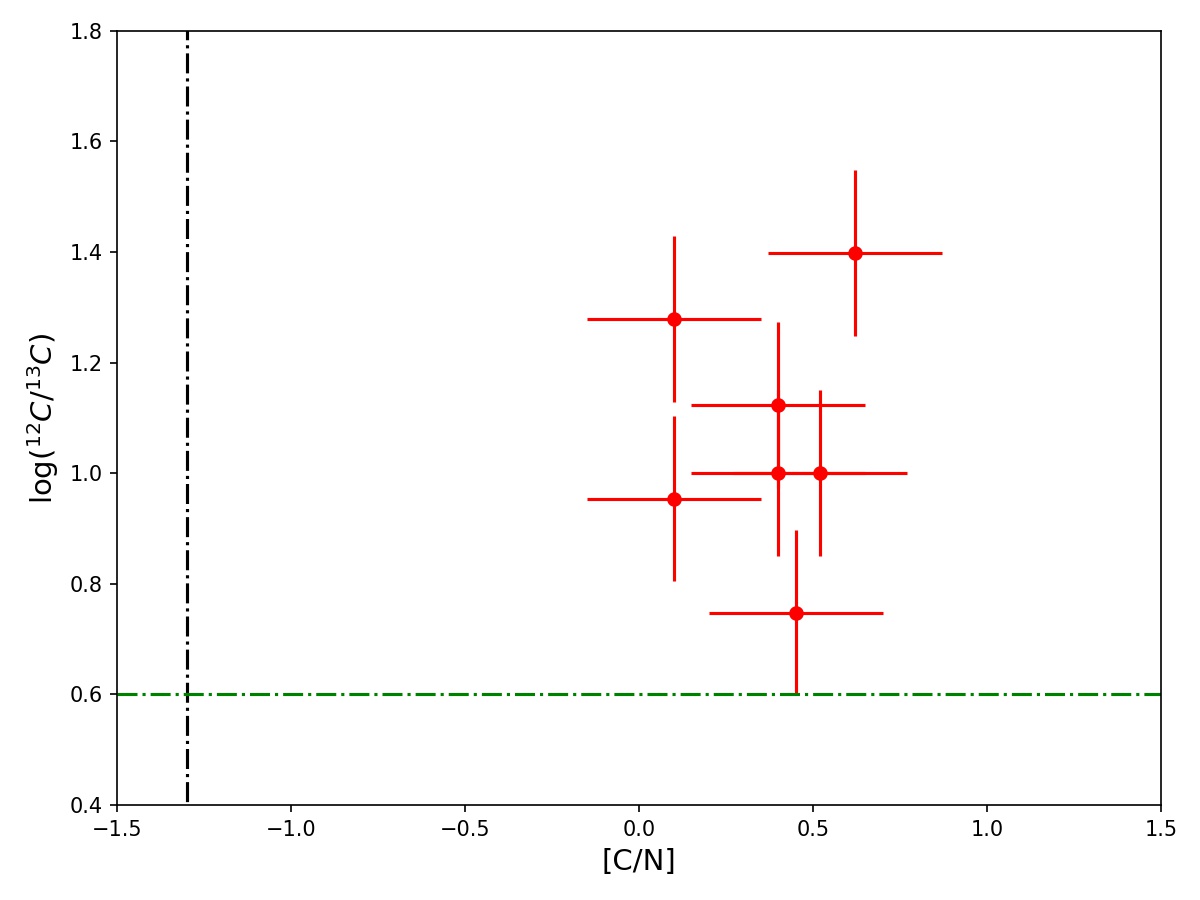}
\caption{The $^{12}$C/$^{13}$C ratio of our program stars are plotted as a function of [C/N]. The dashed lines represent the values for a complete CN cycle where  [C/N] $\sim$ -1.3 and $^{12}$C/$^{13}$C  $\sim$ 4. While the low-mass AGB models (M $<$3.0M$_{\odot}$) predict log($^{12}$C/$^{13}$C) upto 3.4, more massive AGB models predict log($^{12}$C/$^{13}$C) as low as 0.6 due to HBB. Although the low values of $^{12}$C/$^{13}$C of our program stars are similar to the model prediction of HBB, the observed [C/N] values and [hs/ls] ratios of our program stars rule out HBB as a likely scenario for the enrichment of $^{13}$C and N.   So the observed values of $^{12}$C/$^{13}$C and [C/N] could be the resultant of either thermohaline mixing, extra-mixing during first dredge-up,  extra mixing for the RGB bump or the extra mixing in the companion AGB star itself.}

\label{fig:c1213cn}
\end{figure}

The abundance pattern of C, N, O, and Na of our program stars,when compared with the
 models of AGB stars, indicate abundance trend 
with that of a low-mass AGB stars than higher mass AGB stars
(refer figure \ref{fig:cno-comp-agb}).
It can be seen that while the trend of observed abundance matches with that of the models, the observed abundance values are different from the model values. Several factors contribute to this difference including the nucleosynthesis and mixing in the AGB star, the mass of the accreted matter, mixing and dilution of accreted matter on the surface of the CEMP star etc. Since our program stars are all in the giant phase, the accreted matter would have experienced dilution and mixing of processed material from the deeper layers. The observed abundances are not corrected for any such mixing and dilution. The inclusion of mixing and dilution could alter the [C/N] ratios. 
According to  \citet[]{stancliffe2009MNRAS}, the observed C and N abundances among the CEMP giants are sensitive to the mass of the C-rich material accreted from the AGB companion. So the details of accreted mass are required to do the corrections. Hence, a direct comparison of the observed abundances with theoretical models is not performed here to constrain the actual mass of the AGB star instead we restrict our comparison to the trend of abundance distribution.
  The observed Na abundances of the sample is also consistent with low mass AGB contribution (refer figure \ref{fig:nacplusn}).
\\
While there is an overall agreement to the trend of these elements, a  discrepancy is found in the $^{12}$C/$^{13}$C values.
According to low-mass AGB models, the $^{12}$C/$^{13}$C  should be very high upto log ($^{12}$C/$^{13}$C) = 3.4 for AGB stars of mass M $<$ 3M$_{\odot}$ 
\citep{fishlock2014apj} so the observed high $^{12}$C  in the CEMP 
stars is considered to be due to the mass transfer from a low-mass AGB companion.
More massive AGB  models  (M $>$ 3M$_{\odot}$) produce low  carbon isotopic ratio (log ($^{12}$C/$^{13}$C) $<$2.0 ) through Hot Bottom Burning (HBB), a process that happens at the bottom of the convective envelope which converts  carbon to nitrogen when the temperature
is sufficient for CN cycle. The values of carbon isotopic ratio thus obtained
are  similar to the  values observed in CEMP stars. 
However, such  massive companion stars also produce  high nitrogen abundance resulting in [C/N] $<$ 0. HBB also results in the production of lighter $\it {s}$-process elements than heavy $\it {s}$-process elements resulting in [hs/ls] $<$ 0 which is not
found in any of the stars studied here. 
This makes HBB an unlikely scenario for the observed low carbon isotopic ratios.\\
All the program stars in this study are giants
and they would have undergone several mixing episodes during the 
first dredge-up, thermohaline mixing, extra mixing for the RGB bump, etc  \citep[]{charbonnel1998aa,stancliffe2007aa,smiljanic2009aa}.
This changes the surface composition of the carbon and carbon isotopes 
that was originally accreted from the AGB companion and would finally 
result in an intermediate nitrogen enhancement (still [C/N] $>$ 0.1)
 as well as the low isotopic ratio \citep{proffitt1989,barbuy1992aa,keller2001AJ, denissenkov2008apj} (refer figure \ref{fig:c1213cn}).
 While the mixing in the CEMP star is put forward as one of the reasons for the low $^{12}$C/$^{13}$C ratio, the CEMP stars in main-sequence or turn-off stage are also identified to be showing low carbon isotopic ratios \citep{aoki2002apj, lucatello2003aj,masseron2010aa}.  So the measured low $^{12}$C/$^{13}$C ratio could be the resultant of some extra mixing in the companion AGB star itself. \\
The concept of Cool Bottom Processing (CBP) can also be invoked
to explain the moderately high nitrogen abundance 
and low $^{12}$C/$^{13}$C.
In  CBP, the material from the base of the convective envelope 
is mixed into the radiative region located on top 
of the H-burning shell where material captures protons and then
brought back to the envelope thereby leaving CN processed materials
at the surface of the star \citep{wasserburg1995apj}.  Nitrogen and  $^{13}$C are being the byproducts of the CN cycle, a complete
CN cycle produces $^{12}$C/$^{13}$C $\sim$ 4 and [C/N] $\sim -$1.3.
Thus the observed low $^{12}$C/$^{13}$C and [C/N] values 
can be obtained when the carbon is partially processed
possibly in the H-burning shell before reaching the 
convective envelope. Such a mechanism could explain the moderately high nitrogen abundance and low carbon isotopic ratio observed in CEMP stars.  
 CBP is hypothesized to occur in the RGB phase of stars
having M $\leqslant 2.5M_{\odot}$ \citep{charbonnel1995apj}
and in AGB stars where an extra mixing happens at the bottom of the
AGB envelope, and  a part of this envelope material turns carbon
into nitrogen through CN cycle \citep{nollett2003apj, lugaro2017}.
Though the physical mechanism causing the CBP is unclear,
 an extra mixing process is required to explain the observed abundances, 
 which include the instabilities generated due to the 
molecular weight inversion induced by $^3$He burning \citep{eggleton2008apj}
 and the magnetic buoyancy induced by a stellar dynamo 
 \citep{busso2007apj, palmerini2008}. \\
For the case of oxygen abundance, low-metallicity AGB stars are expected to produce an elevated
oxygen abundance from $^{12}$C($\alpha$, $\gamma$)$^{16}$O reaction 
activated due to the hotter conditions in the helium burning shell. 
But the oxygen yields predicted by different models do not agree with 
each other. \citet{herwig2004apjs} find that the oxygen abundance 
decreases with mass of the AGB star due to the dilution in the AGB 
envelope. Though the dredge up matter brings more oxygen to the surface, 
the dilution dominates the dredge up when the mass increases. 
\citet{karakas2007pasa} found a contradicting scenario. 
According to their models, as the mass increases, in the He shell, the reaction $^{12}$C($\alpha$, $\gamma$)$^{16}$O is favored by the competition between the dilution and the hotter conditions  and around 3M$_{\odot}$, 
maximum oxygen yield is predicted. 
All the models assume a solar scaled abundance for the key 
elements to start with. The amount of the elements produced
during AGB evolution is so high that the initial assumptions do
not affect the final composition. In contrary,   the oxygen is produced in the scale of [O/Fe]$\sim$0.5
in metal-poor stars \citep{lucatello2011apj} which is
greatly influenced by the initial composition. 
 \\
Another scenario which can account for observed CNO abundance, especially 
 the low carbon isotopic ratio
and moderately high nitrogen abundance, is the Proton Ingestion Episodes
(PIEs). At the beginning of the thermally pulsating phase of low-mass low-metallicity AGB stars, the convective envelope
extends up to the base of the H-rich envelope and the
protons get mixed in to the convective shell which are captured by  $^{12}$C. This produces $^{13}$C and  $^{14}$N  through various reactions
which results in a deep-Third Dredge Up (TDU). After the deep TDU,
 the model follows the standard AGB evolution. Each thermal pulses 
bring the synthesized material to the surface of the stellar envelope
resulting in constantly varying the surface composition. The observed
high  $^{13}$C is also produced in the PIE. The final abundance
pattern at each TDU phases resembles to the CNO abundance pattern
exhibited by the stars in this study (refer figures 6 and 7 in \citet{cristallo2009pasa}. 
Regardless of the abundance pattern exhibited, certain uncertainties 
are also associated with the PIEs.
Enhancement in the alpha elements can suppress the PIEs as the proton
ingestion requires lower CNO abundance to break the entropy barrier. 
Also the mixing and nucleosynthetic mechanisms in the convective
regions do not account for the assumptions of mixing length theory 
and one dimensional spherically symmetric evolution \citep{cristallo2009pasa, herwig2011apj}.
Nevertheless, we can associate the PIEs with a few TDUs  contributing to the observed 
CNO abundances and low isotopic ratio in CEMP stars. 
\subsection{Neutron capture elements to constrain the nucleosyntheses in the companion AGB star}
\begin{figure*}
\centering
\includegraphics[width = \textwidth]{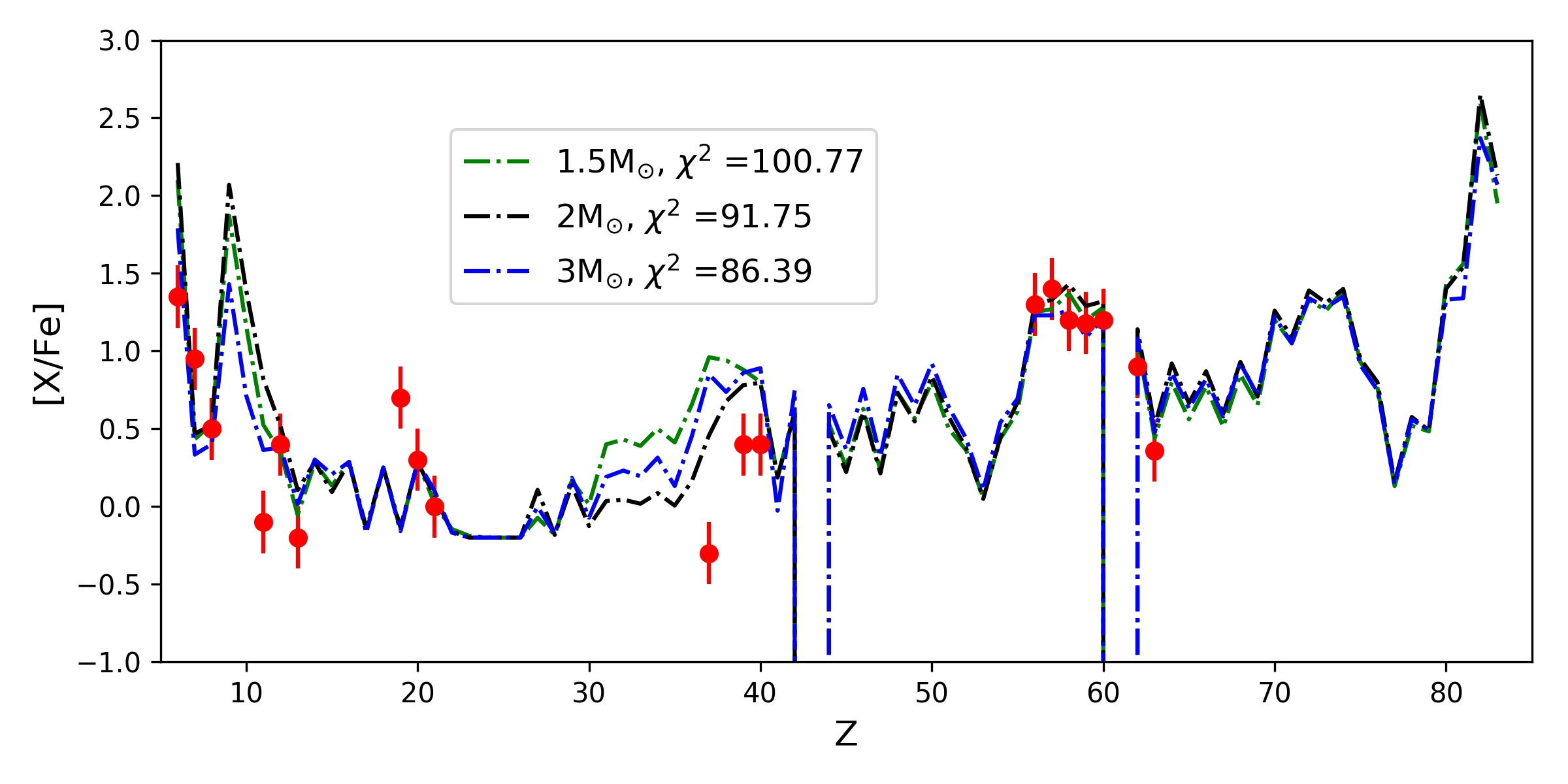}
\caption{Various elemental abundances of the \cemps\ star, HE 1152-0355 (red filled circles),  are fitted with AGB models of different masses. The abundance pattern for different masses and the respective chi-square values are listed in the legend}.
\label{fig:he11}
\end{figure*}
\begin{figure}
\centering
\includegraphics[width = 9.0cm, height= 7cm]{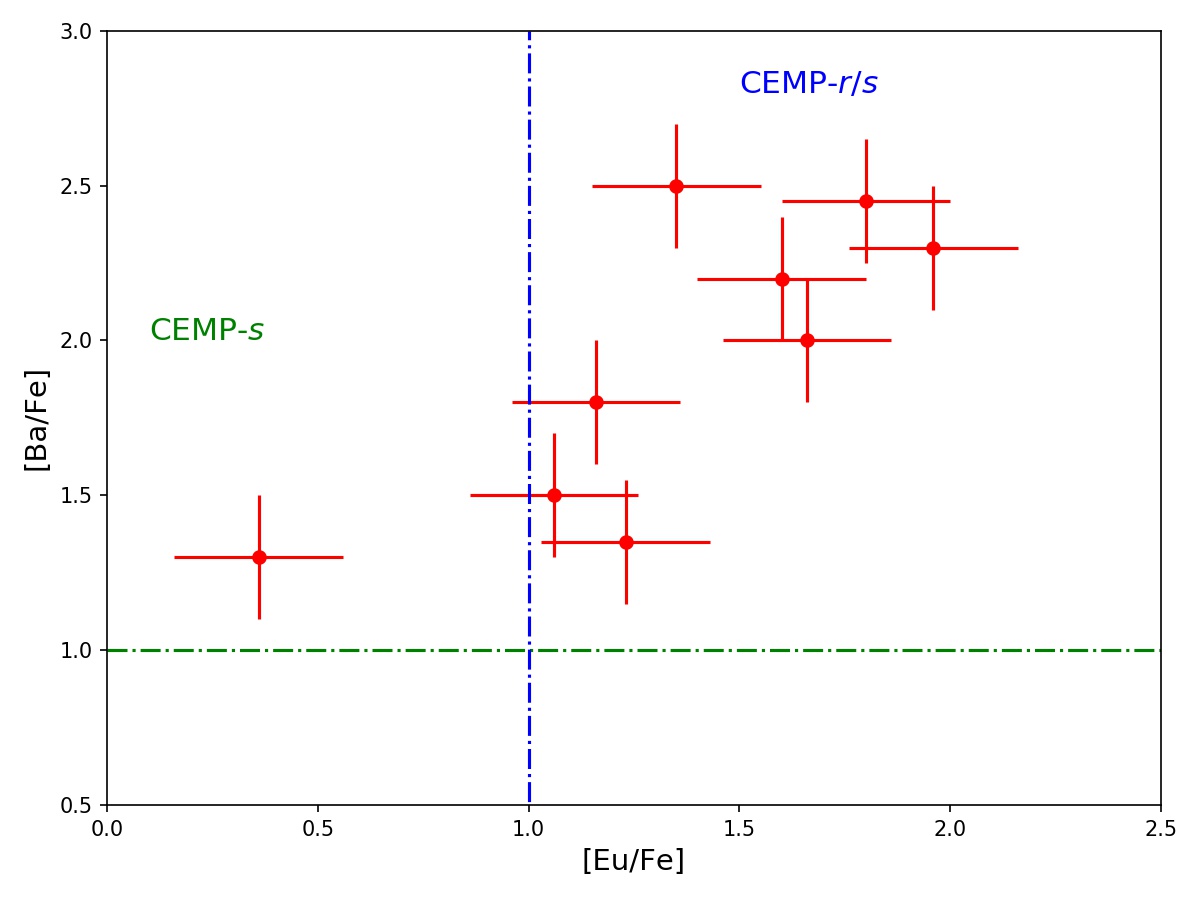}
\caption{The [Ba/Fe] and [Eu/Fe] of all program stars are plotted here. The region of various CEMP population is marked in the figure. The region where [Ba/Fe] $>$ 1.0 and [Eu/Fe] $>$ 1.0 is where the \cemprs\ population occupies whereas the CEMP- $\it{s}$ population fills the region where [Ba/Fe] $>$ 1.0 and [Eu/Fe] $<$ 1.0.}
\label{fig:euba}
\end{figure}

All the program stars in this study are rich in both $\it {s}$-process and $\it {r}$-process elements. Among the program stars, HE 1152-0355 satisfies the criteria to be classified  as a \cemps\ star. So the abundance pattern of this star is compared with that from low-mass AGB models. We compared the AGB models from F.R.U.T.Y   (Full-network repository of Updated Isotopic Tables \& Yields) database \citep{cristallo2011ApJS} where we chose models comparable to the metallicity of HE 1152-0355.  [Rb/Zr]  being an important tracer for the mass of the companion AGB stars in \cemps\ population, we use the [Rb/Zr] ratio of the star to constrain the masses of the models.
Since the star exhibits a low [Rb/Zr] ratio ([Rb/Zr] $<$ -0.7), we restricted the comparison to only low-mass models which is also consistent with the results from the section \ref{cnoagbmass}.   The fitting is done with the available models (1.5M$_{\odot}$, 2M$_{\odot}$, 3M$_{\odot}$) in the database and is shown in the  figure \ref{fig:he11}. All these  models exhibit similar  distribution for most of the elements and the abundances of $\it{hs}$-elements in HE 1152-0355  matches with all the model values within the uncertainty. Since we do not have the information of mixing and dilution on the surface of the CEMP star and the efficiency of mass-transfer from the companion AGB star, we restrict our analysis to comparing the model values with the observed ones instead of addressing the reasons behind the difference in various abundances with the model predicted values.  
Except HE 1152-0355, all the stars have [Ba/Fe] $>$ 1.0, [Eu/Fe] $>$ 1.0  and  [Ba/Eu] $>$ 0 (refer figure \ref{fig:euba}).
Thus we classify these stars as \cemprs. (refer table \ref{tab:stparam}). 

Various theories have been put forward to explain the origin of both $\it {s}$- and $\it {r}$-process elements in \cemprs\ stars.
 Eu being the r-process representative \citep{beers-christlieb2005ARA&A}, 
its enhancement  along with the enhancement in Ba, the 
$\it {s}$-process representative, cannot be explained with standard $\it {s}$-process in 
AGB stars \citep{lugaro2012apj}. 
\begin{figure*}
\centering
\begin{tabular}{c c}
\includegraphics[width=0.45\textwidth]{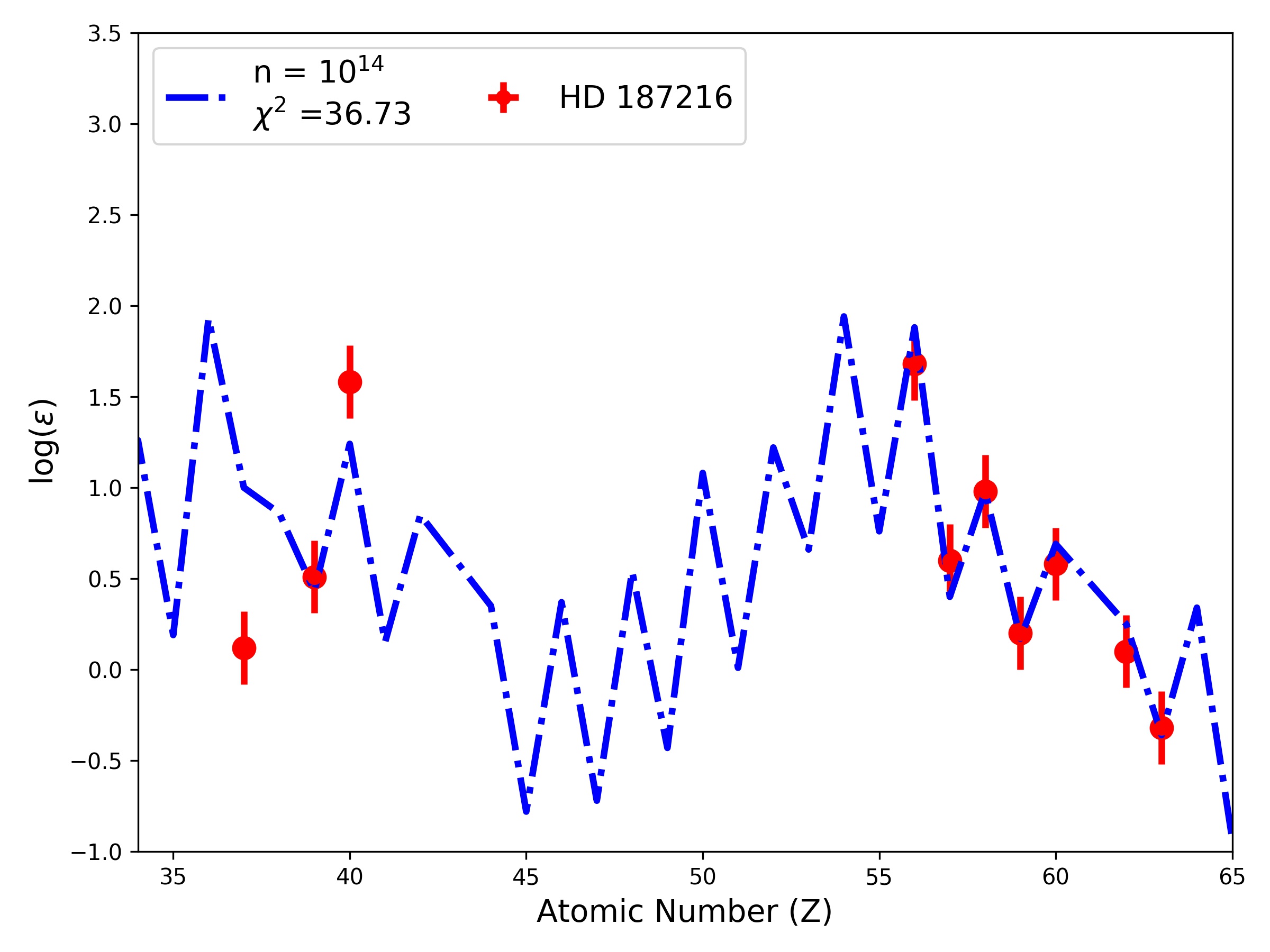} &
\includegraphics[width=0.45\textwidth]{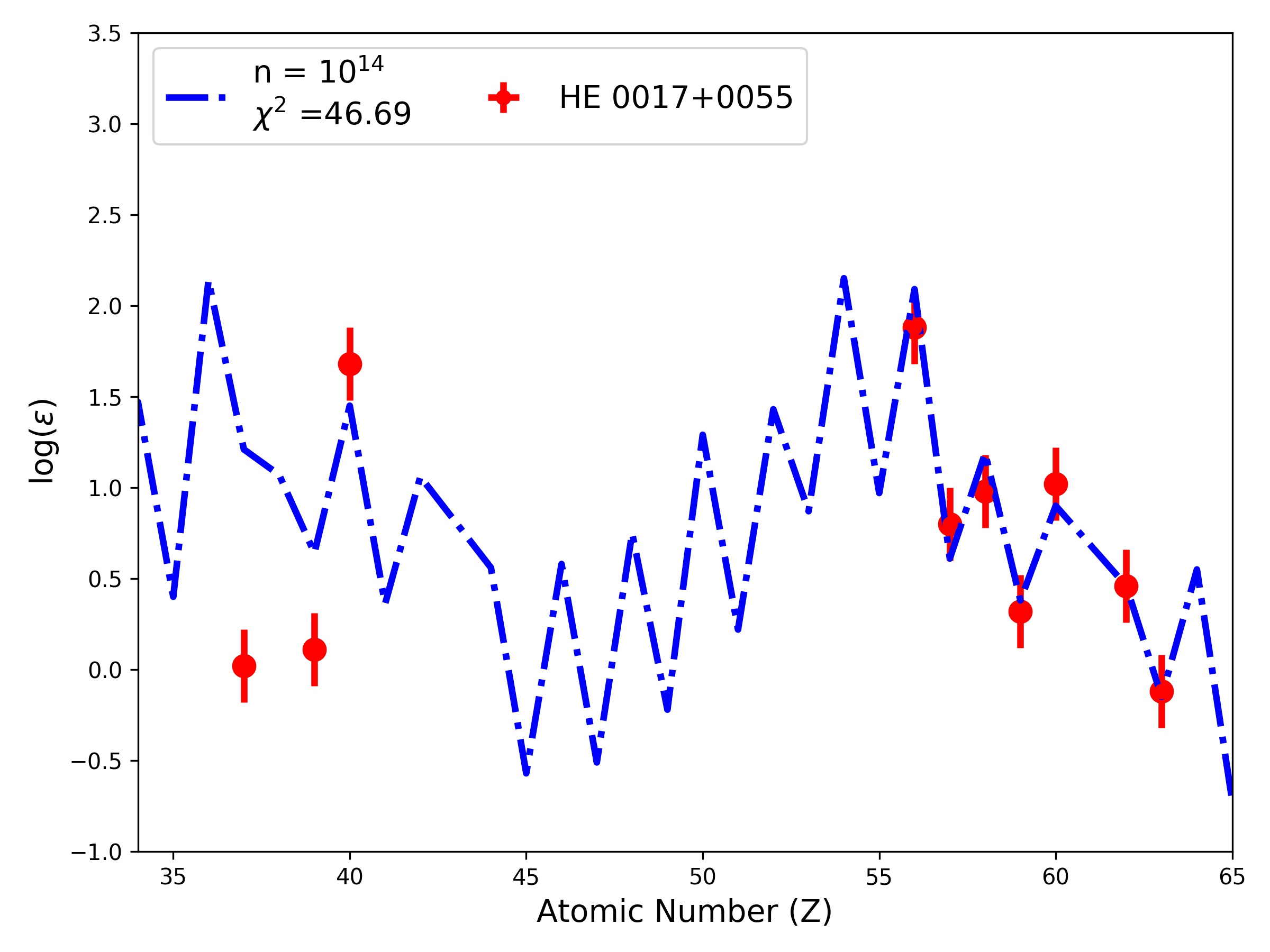} \\
\includegraphics[width=0.45\textwidth]{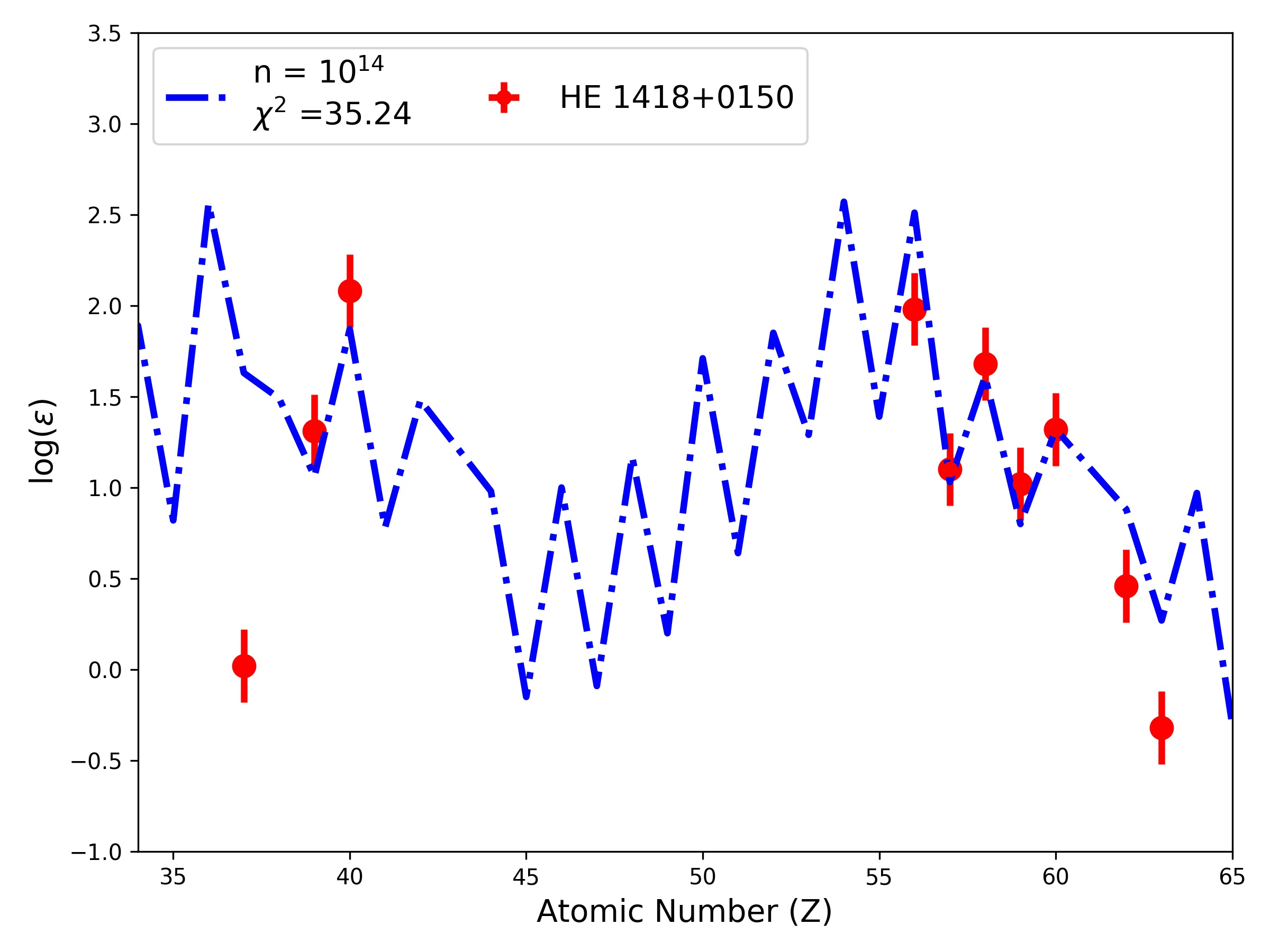} &
\includegraphics[width=0.45\textwidth]{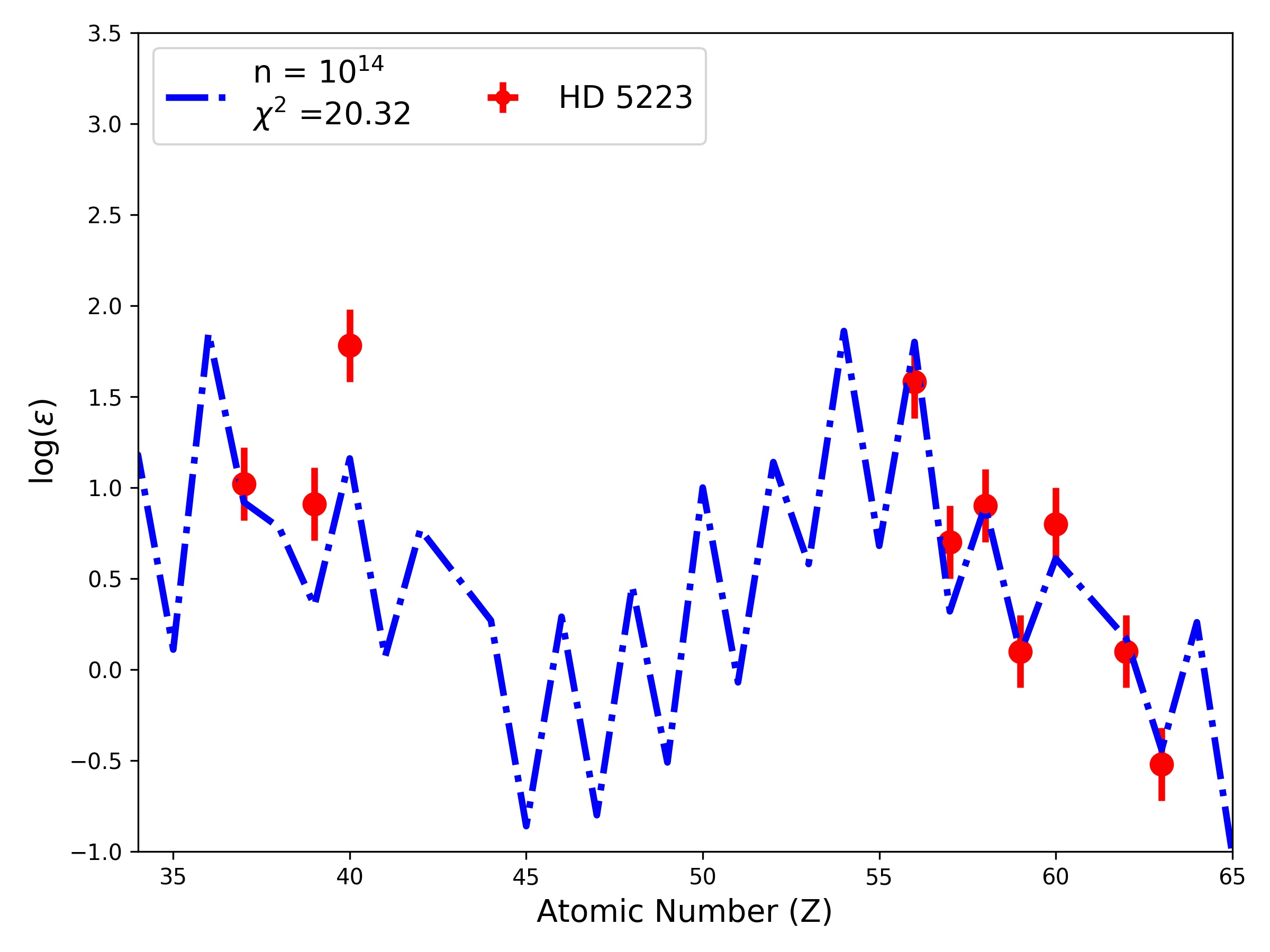} \\
\includegraphics[width=0.45\textwidth]{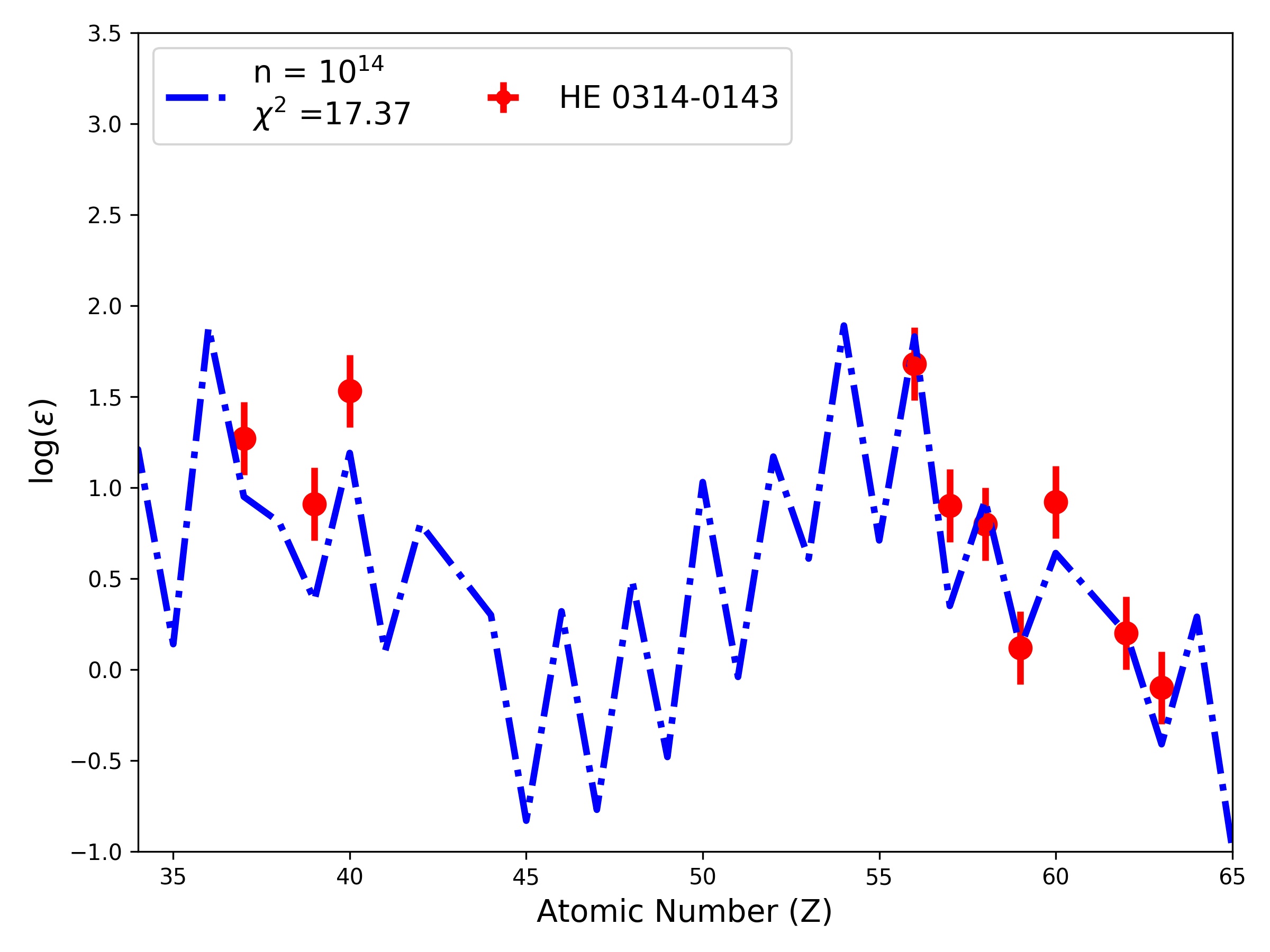} &
\includegraphics[width=0.45\textwidth]{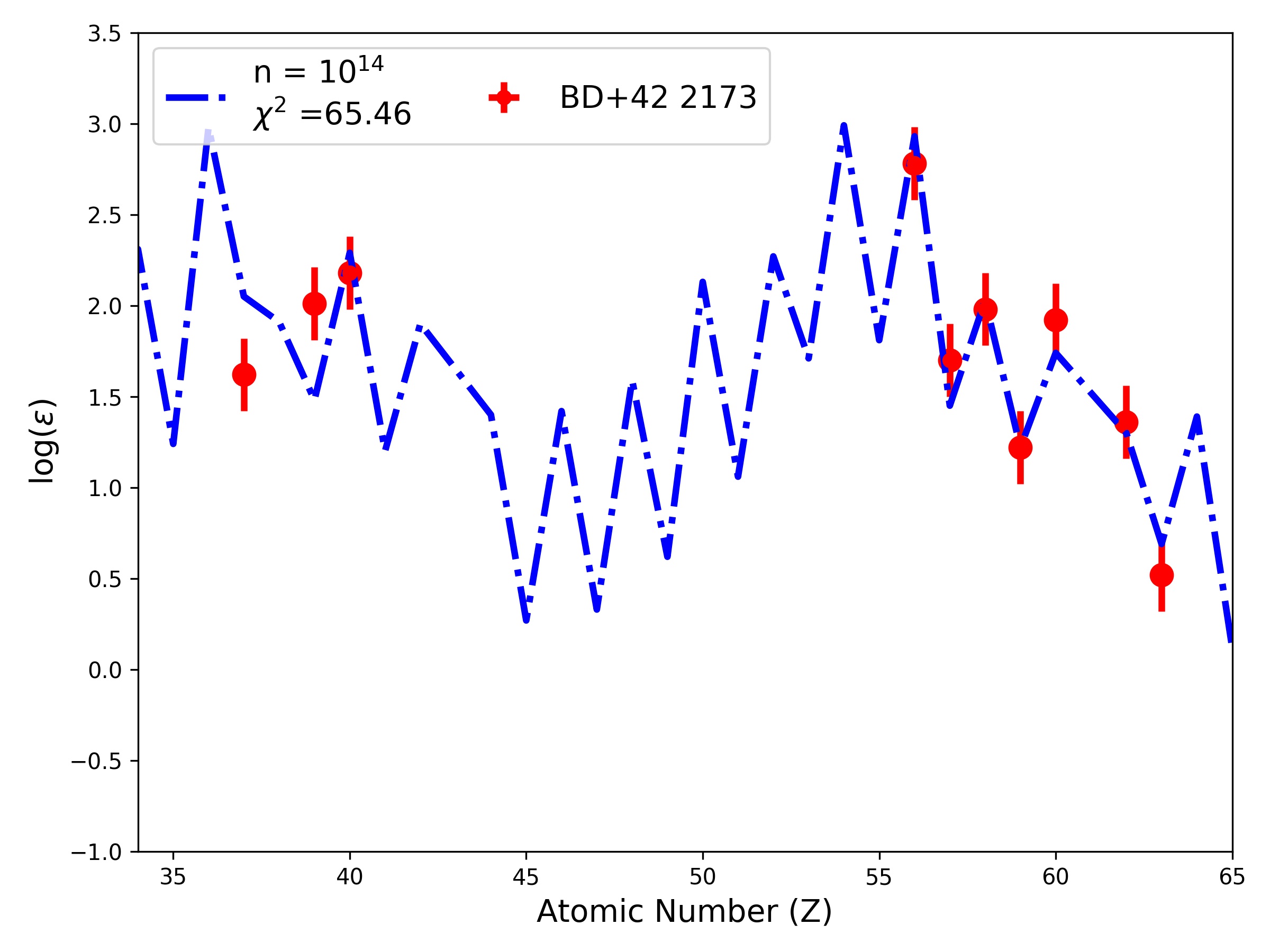} \\
\end{tabular}
\caption{The CEMP-$\it {i}$ models from \citet{hampel2016apj} is fitted with the heavy elemental abundance of stars from this study.}
\label{fig:cempi}
\end{figure*}
While the $\it {s}$-process requires 
neutron densities, $\it{n}$ $\approx$ 10$^{6}$ - 10$^{10}$ cm$^{-3}$ \citep{busso1999araa}, the  r-process requires
$\it{n}$ higher than 10$^{20}$ cm$^{-3}$  \citep{goriely2005nupha,thielemann2011}. So different formation sites have been proposed to explain the origin of these elements. In which, the carbon and $\it{s}$-process  elements originate from the AGB companion similar to that in \cemps\ stars where the birth cloud  of the binary system was previously enriched with the r-process elements from supernovae explosions  and/or neutron star mergers \citep{bisterzo2012MNRAS}.

Recent computations explored an  intermediate neutron densities to $\it{s}$- and $\it{r}$-processes and they were successful in producing the abundance patterns of  \cemprs\ stars \citep{hampel2016apj, denissenkov2017apj}. The neutron-capture process in this intermediate neutron density ($\sim$ 10$^{15}$ cm$^{-3}$) is called as \iprocess\ and is activated by the rapid ingestion
of a significant amount of H in He-burning convective regions \citep{cowan1977apj}. 
 Even though the nucleosynthetic sites for \iprocess\ is still a debate, the proposed sites  include: low-mass, low-metallicity AGB stars \citep{campbell2008aa, campbell2010aa, cruz2013aa, cristallo2016apj}, super-AGB stars \citep{doherty2015, jones2016},  evolved low-mass stars \citep{herwig2011apj, hampel2016apj, hampel2019ApJ}, rapidly accreting white dwarfs (RAWD) in closed binary systems \citep{herwig2014ApJ, denissenkov2017apj} and massive metal-poor stars \citep{banerjee2018ApJ, clarkson2018MNRAS}. \\
We compared the neutron-capture elemental abundance pattern of our program stars with available \iprocess\ models of similar metallicity \citep{hampel2016apj, denissenkov2019}.
The figure 11  and 12  in \citet{denissenkov2019} predicts over enhancement for neutron-capture elements than the observationally  derived values for our program stars. So the abundances are not matching with the RAWD models values \citep{denissenkov2019} whereas for the case of \citet{hampel2016apj} models, we find a good agreement (refer figure \ref{fig:cempi}). 
As explained in the sec \ref{cnoagbmass}, at the bottom of the convective envelope, the $^{13}$C produced in PIEs acts as a source of neutrons through $^{13}$C($\alpha$, n)$^{16}$O reaction. These PIEs can produce necessary neutron densities as large as 10$^{15}$ cm$^{-3}$ (\citet{cristallo2009pasa}). 
 We have only considered elements 37 $\le$ Z $\le$ 63 since the model is available only for elements heavier than Fe.
We used the equation given in \citet{hampel2016apj} to calculate the model abundance on the surface of the \cemprs\ stars:
$$X = X_{i}*(1-d)+X_{\odot}*d$$
where $X_{i}$ is the calculated \iprocess\ abundance of each element from the model, $X_{\odot}$ is the scaled-solar abundance, and {\rm d} the dilution factor, which provides the measure of how much  $\it{i}$-processed material is mixed with unprocessed material. However, the dilution factor in this equation does not provide any quantitative measure on various specific mixing mechanisms on the AGB surface or mass-transfer  onto the CEMP star \citep{hampel2019ApJ}, instead it is a free parameter and is varied until we obtain a best-fitting  for a model and the observed abundances for different neutron densities. Since the dilution factor does not have any physical significance, we are not quoting the value. The best-fit is calculated using the equation:
$$\chi^{^2} = \sum_{Z=37}^{63}\frac{[X_{Z}/Fe]_{obs} - [X_{Z}/Fe]_{mod}}{\sigma^{2}_{Z,obs}}$$
Where $\sigma^{2}_{Z,obs}$ is the observational error of $[X_{Z}/Fe]_{obs}$ and  $[X_{Z}/Fe]_{obs}$ and $[X_{Z}/Fe]_{mod}$ are observed and model abundances of the element with atomic number Z. $\chi^{^2}$ indicates the goodness of the fit to  tell which model matches  the observations
best, but its value does not have the statistical meaning in the
conventional sense. The best fit model parameters and the corresponding $\chi^{^2}$  is mentioned in each plot for all the program stars (refer figure \ref{fig:cempi}). \\
\\
The models from \citet{hampel2016apj} provide a fairly good match with the observed data. However, the authors have noted that the neutron-capture processes considered in \citet{hampel2016apj} use one-dimensional single-zone nuclear-network calculations   without clearly distinguishing the stellar sites hosting such a process. So, in realistic scenarios, the heavy elements produced may demonstrate a different abundance pattern which cannot be clearly stated as there are no evidences to conclude so. 
\citet{hampel2016apj} also claim that the light-elements are not  produced by a significant amount in the neutron-capture processes, but large production of  $^{13}$C and $^{14}$N in PIEs are illustrated \citep{cristallo2009pasa, cristallo2016apj}. So, multidimensional simulations are required for the \iprocess\ nucleosynthetic processes in order to accurately  model the abundances of these light elements and their isotopic ratios  \citep{herwig2011apj, stancliffe2011ApJ, herwig2014ApJ}. 
Also, the effect of multiple PIEs to the enrichment of $\it{ls}$ and $\it {hs}$ are not available with current models.  While the shorter neutron bursts can produce lighter s-peak elements, the longer neutron exposures produce the heavier s-peak elements  \citep{koch2019aa}. In the figure \ref{fig:cempi}, the light $\it {s}$-process elements, Y and Zr are not well-fitted, so a combination of the individual neutron burst events may be required to model the lighter $\it {s}$-process elements. Moreover, the number of neutron-capture elements detected in the spectra of our program stars  are rather small.  So it is not very clear that the heavy element enrichment in our program stars are produced only through a pure i-process or a mixture of other neutron-capture processes along with the i-process. 
\section{Conclusions}
We studied 7 CEMP stars using both high-resolution optical spectra and low-resolution NIR spectra. We derived the O and $^{12}$C /$^{13}$C ratio using optical and NIR spectra and found good  agreement in the derived abundances. All the CEMP stars  in this study exhibit enhancement in both carbon and neutron-capture elements and also show radial velocity variations. This favors 
AGB mass transfer scenario as a possible mechanism for the enhanced carbon and n-capture elements.
Both the light element and the n-capture elements indicate accretion from a low-mass AGB companion. The abundance pattern of C, N and O is compared with AGB models of different masses and all the stars in this study are found to have low-mass low-metallicity AGB stars as companions,  This low-mass nature of the companion is further confirmed  by the Na abundances. 
\\
We demonstrated that low-resolution NIR observations can complement high resolution optical observations to derive the oxygen abundance and $^{12}$C /$^{13}$C ratio  in cool CEMP stars. Since the abundance of C, N can also be derived using low-resolution spectra, it may be proposed that low-resolution optical and NIR observations are adequate to constrain the nature of binary companion in cool CEMP stars. This will enable us to probe fainter cool CEMP stars thereby probe several cool CEMP stars beyond Milky Way \\
All the 7 CEMP stars studied here are identified to be exhibiting enhancements in both Ba and Eu. Among them, one of the stars (HE 1152-0355) exhibits [Ba/Fe] $>$ 1.0, [Ba/Eu] $>$0.0 and [Eu/Fe] $<$ 1.0.  These values  sub-classify the star as a \cemps\ category. The low-mass AGB models fit well with the observed abundance pattern of HE 1152-0355.  Rest of the 6 stars exhibit [Ba/Fe] $>$ 1.0, [Ba/Eu] $>$0.0 and [Eu/Fe] $>$ 1.0. This indicates that these stars are enhanced in both $\it {s}$-process and $\it {r}$-process elements which sub-classify them as \cemprs. Among the 6 stars, 5 stars are classified as \cemprs\ in this study using their [Eu/Fe] abundances which were not available earlier in literature thus they were wrongly classified as \cemps\ stars. The abundance pattern of the neutron-capture elements in these \cemprs\ stars when compared with the models of \iprocess\ from \citet{hampel2016apj} provides a good match. This support the idea of the operation of \iprocess\ in the inter-shell region of low-mass low-metallicity AGB stars for the production of heavy elements seen in the \cemprs\  stars.  While the \citet{hampel2016apj} models provide a fairly good match in the fitting with the observed abundances, the light s-peak elements do not match well with the model as compared to the heavy s-peak elements. This may demand the inclusion of multiple PIEs of different neutron burst events  which is currently not available with the existing models.
Also, the number of neutron-capture elements detected in the spectra of program stars are somewhat smaller than  required to differentiate between a pure  \iprocess\ model and a mixture of neutron-capture events that would have contributed to the resultant abundance pattern.
Even though the low-resolution optical and NIR spectroscopy is sufficient to constrain the mass and metallicity of the companion AGB star of cool CEMP stars, high-resolution studies can  provide the better understanding of the nucleosynthetic origin of various elements present in them. Combining the key elements such as Li, C, N, O abundances and n-capture elements will provide valuable insights to various mixing mechanism and nuclear process that might have contributed to the observed abundances in \cemprs\ stars.
\section*{Acknowledgements}
We thank the anonymous referee for the useful comments.
We thank the staff of IAO, Hanle and CREST,
Hosakote, that made these observations possible. The facilities at IAO and CREST are operated by the Indian Institute of Astrophysics, Bangalore. AS and DKO acknowledge the support of the Department of Atomic Energy, Government of India, under Project Identification No. RTI 4002.  AS thanks Drisya Karinkuzhi for the fruitful discussions on different AGB models and \iprocess\ contributions to the enhancement  of neutron-capture elements. 
\section*{Data availability}
The data underlying this article will be shared on reasonable request to the corresponding author.



\bibliographystyle{mnras}
\bibliography{reference} 
\appendix

\onecolumn
\section{Additional table of spectral lines used for abundance analysis}
\begin{longtable}[c]{cccc}
\label{tab:linelist}
\begin{tabular}{|c|c|c|c|}
\hline

Element & $\lambda$ & $\chi$ & log ${\it gf}$ \\
\hline
\hline
O I & 6300.30 & 0.00 & -9.72 \\
Na I & 5682.63 & 2.10 & -0.71 \\
Na I & 5688.21 & 2.10 & -0.45 \\
Mg I & 5528.41 & 4.35 & -0.62 \\
Mg I & 5711.09 & 4.35 & -1.83 \\
Al I & 6696.02 & 3.14 &  -1.35 \\
Al I & 6698.67 & 3.14 & -1.65  \\
K I & 7664.91 & 0.00  & 0.13  \\
K I & 7698.97 & 0.00  & -0.17  \\
Ca I &5588.75 &2.53 &0.34 \\
Ca I &6102.72 &1.88 &-0.79 \\
Ca I &6122.22 &1.89 &-0.32 \\
Ca I &6162.17 &1.90 &-0.09 \\
Sc II & 5526.79 & 1.77 &  0.02 \\
Sc II & 5658.36 & 1.50 & -1.21 \\
Sc II & 5684.20 & 1.51 & -1.07 \\
Rb I & 7800.26 & 0.00 &  0.14 \\ 
Y II & 5200.41 & 0.99 & -0.57 \\
Y II & 5205.73 & 1.03 & -0.34 \\
Zr II & 6127.48 & 0.15 & -1.06 \\
Zr II & 6134.59 & 0.00 & -1.28  \\
Ba II & 5853.68 & 0.60  & -0.91  \\
Ba II & 6141.73 & 0.70  & -0.03  \\
La II & 4920.97 & 0.13 & -2.26 \\
La II & 5290.82 &  0.00 & -1.65 \\
La II & 5301.97 & 0.40 &  -2.59 \\
La II & 5303.51 & 0.32 & -1.87 \\
La II & 5797.57 & 0.24 & -1.38 \\      
La II & 5805.77 & 0.13 & -1.59 \\       
Ce II & 5187.46 & 1.21  & 0.30 \\
Ce II & 5274.23 & 1.04 & 0.13 \\
Ce II & 5330.56 &  0.87 & -0.40 \\
Pr II & 5206.56 & 0.95 & -0.16 \\
Pr II & 5207.90 & 0.80 & -0.58 \\
Pr II & 5219.05 & 0.80 &-0.05 \\
Pr II & 5259.73 & 0.63 & 0.11 \\
Nd II & 4811.34 & 0.06 & -1.02 \\
Nd II & 5215.65 & 1.26 & -0.65 \\
Nd II & 5249.58 &  0.98 &  0.09 \\
Nd II & 5276.87  & 0.86  & -0.39 \\
Nd II & 5319.81 & 0.55 & -0.14 \\
Sm II & 4777.84 & 0.04 & -1.28 \\
Sm II & 4791.58 & 0.10 & -1.24 \\
Sm II & 4815.80 & 0.19 & -0.82 \\
Sm II & 4948.63 & 0.54 & -0.85  \\
Sm II & 4952.37 & 0.33 & -1.37  \\               
Sm II &  5202.70 & 0.33  & -1.65 \\
Eu II & 6645.09 & 1.38  & 0.12 \\
\hline
\end{tabular}
\end{longtable}
\section{Additional figures of spectral fitting}
\begin{figure}
    \centering
    \includegraphics[width = \textwidth]{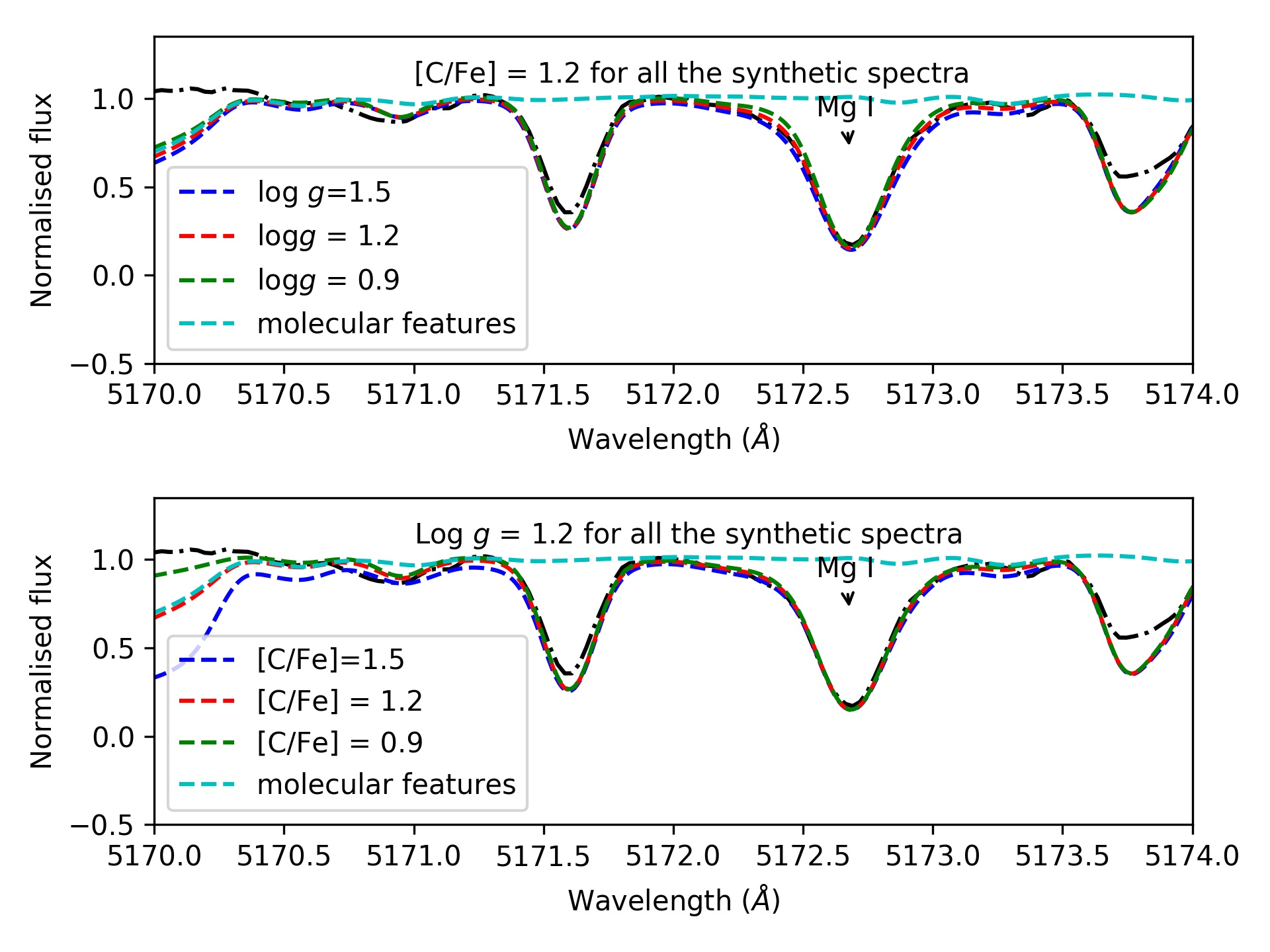}
    \caption{The effect of \logg\ and Carbon abundance on Mg I lines are plotted. The top panel in the figure is the fitting of Mg I line at 5172 \AA\ with different values of \logg\ but with a fixed C abundance to show the sensitivity of \logg\ to the Mg line wings. The bottom panel in the figure shows fitting of Mg I lines at 5172 \AA\  with synthetic spectra of different values of C abundance but with a fixed value of \logg. It can be seen that, the C abundance does not have much effect on these Mg I lines.}
    \label{fig:c2mg5172}
\end{figure}

\bsp	
\label{lastpage}
\end{document}